\documentclass{aa} 
\usepackage{natbib, my_def, graphicx} 
\bibpunct{(}{)}{;}{a}{}{,}
\bibstyle{aa}
\def\udier{\"u}

\def \etal   {{~et~al.~}}
\begin{document}
\title{Radio Polarimetry of 3C\,119, 3C\,318, and 3C\,343 at milliarcsecond 
resolution}
 
\author{ F. Mantovani \inst{1} \and
         A. Rossetti  \inst{1} \and
	 W. Junor     \inst{2} \and
	 D.J. Saikia  \inst{3,4} \and
	 C.J. Salter  \inst{5} 
         }

\offprints{Franco Mantovani\\
  \email{fmantovani@ira.inaf.it}}

\institute{Istituto di Radioastronomia -- INAF, via Gobetti 101,
 I--40129, Bologna, Italy 
 \and Los Alamos National Laboratory, Los Alamos, NM 87545, USA
 \and National Centre for Radio Astrophysics, TIFR, Post Bag 3,
 Ganeshkhind, Pune 411 007, India
 \and ICRAR, University of Western Australia, Crawley, WA 6009, Australia
 \and Arecibo Observatory, HC3 Box 53995, Arecibo, Puerto Rico 00612
}

\date{Received \today; accepted ???}

\abstract
{}
{We report new Very Long Baseline Array (VLBA) polarimetric observations of the Compact
Steep-Spectrum (CSS) sources 3C\,119, 3C\,318, and 3C\,343 at 5 and 8.4~GHz. }
{By using multifrequency VLBA observations we have
derived milliarcsecond-resolution images of the total intensity,
polarisation, and rotation measure ($RM$) distributions.
}
{ The CSS source 3C\,119, associated with a possible quasar, has
source rest-frame $RM$ values up to $\sim$10200 rad m$^{-2}$  in a
region which coincides with a change in the direction of the inner
jet.  This component is located $\sim$325 pc from the core, which is
variable and has a peaked radio spectrum.  In the case of 3C\,318,
associated with a galaxy, a rest-frame $RM$ of $\sim$3030 rad
m$^{-2}$ has been estimated for the brightest component which
contributes almost all of the polarised emission. Further, two more
extended components have been detected, clearly showing ``wiggles''
in the jet towards the southern side of the source. The CSS source
3C\,343 contains two peaks of emission and a curved jet embedded in
more diffuse emission. It exhibits complex field directions near the
emission peaks, which indicate rest-frame $RM$ values in excess of
$\approx$6000 rad m$^{-2}$.  The locations of the cores in 3C\,318
and 3C\,343 are not clear.
}
{The available data on mas-scale rest-frame $RM$ estimates for CSS
sources show that these have a wide range of values extending up
to $\sim$40000 rad m$^{-2}$ in the central region of OQ172, and
could be located at projected distances from the core of up to
$\sim$1600 pc, as in 3C\,43 where this feature has a rest-frame $RM$ of
$\sim$14000 rad m$^{-2}$.  $RM$ estimates for cores in core-dominated
radio sources indicate that in addition to responding to an overall
density gradient of the magneto-ionic medium, geometry, orientation
and modes of fuelling
may also play a significant role. In addition to these effects, the
high values of $RM$ in CSS sources are possibly due to dense clouds of
gas interacting with the radio jets. The observed distortions in the
radio structures of many CSS sources are consistent with this
interpretation.  }


\titlerunning{Radio Polarimetry of 3C\,119, 3C\,318, 3C\,343}
\maketitle

\maketitle 
\section{Introduction} 
The number of Compact Steep-Spectrum (CSS) sources with detailed
polarimetric information available at milliarcsecond resolution is
still small. Polarised radio emission from CSS radio galaxies is
either very weak or below the detection limits at centimetre
wavelengths. In contrast, CSS quasars show linear polarisation
percentages of up to 10\% above 1\,GHz \citep [ ] [and references
therein] {Rossetti08}.  We have conducted a series of observations of
CSS sources having significantly polarised emission and high values of
rotation measure ({\it RM}) using the Very Long Baseline Array (VLBA).

CSS objects are {\it young} radio sources, with ages $<10^{{\rm 3-5}}$\,yr.
They have linear sizes $\leq 20$\,kpc
\footnote {$H_0=71\,{\rm km}\, {\rm s}^{-1}\, {\rm Mpc}^{-1}, \Omega_{\rm m}=0.27, \Omega_{\rm vac}=0.73$} 
and steep high-frequency radio spectra ($\alpha >0.5$; ${\rm
S}_{\nu}\propto\nu^{-\alpha}$). Being sub-galactic in size, CSS sources
reside deep within their host galaxies.  Therefore, Faraday
rotation effects are to be expected when their polarised synchrotron
emission is observed through the host galaxy magneto-ionic
interstellar medium (ISM). The comparison of polarised emission
over a range of wavelengths is an important diagnostic of the physical
conditions within and around these compact radio sources (see Cotton
et al. 2003c for an overview).

Existing sub-arcsec polarimetry has provided evidence in favour of
the interaction of components of CSSs with dense clouds of gas, as
seen for example in the CSS quasar 3C\,147 \citep {Junor99a}.

Results for the first two CSS quasars observed in our on-going
program, \object{B$0548+165$} and \object{B$1524-136$}, are available
in  \citet{Mantovani02}, while those for \object{3C\,43}
(B$0127+233$) are to be found in \citet{Mantovani03}. More
recently, the results for \object{3C\,147} (\object{B$0538+498$})
have been presented by ~\citet{Rossetti09}.  These sources have all
been imaged with milliarcsecond resolution by means of full-Stokes
VLBA observations.

In this paper, we report on multi-frequency VLBA, plus a single Very
Large Array (VLA) antenna, polarisation observations at 5 and
8.4\,GHz for \object{3C\,119} (\object{B$0429+415$}), \object{3C\,318}
(\object{B$1517+204$}), and \object{3C\,343} (\object{B$1634+628$}).

In Section \ref{sec:observation} we summarise the observations and data
processing. Section \ref{sec:sources} describes the new information
obtained on the structural and polarisation properties of
\object{3C\,119, 3C\,318, and 3C\,343}. Discussion and conclusions are
presented in Sections \ref{sec:discussion}  and \ref{sec:conclusions}
respectively.

\section{Observations and data reduction} 
\label{sec:observation}


Polarimetric observations of \object{3C\,119}, \object{3C\,318}, and
\object{3C\,343} using the VLBA and one VLA antenna were carried out
at 5 and 8.4\,GHz as detailed in Table~\ref{tab:parms}.  The
data were recorded in both right- and left-circular polarisation in
four 8-MHz bands. At 5\,GHz these bands were spread across the
available bandwidth of $\approx 500$~MHz, allowing us to obtain truly
simultaneous, independent, polarisation images. Only two of the
four sub-bands could  make use of the VLA antenna in the array due to
limitations in the available VLA 5\,GHz system.  In order to
increase the sensitivity to polarised emission at 8.4~GHz, we chose
to use contiguous IFs for this band.
\begin{table*}[htbp]
\centering
 \caption{VLBA+VLA1 observing parameters.}
 \label{tab:parms}
\tabcolsep 1.4mm
 \begin{tabular}{cccccc}
 \hline
 \hline
Sources   & Obs. Date   & Duration  & Fringe Finder      & Polarisation Calibrators \\
\hline 
 3C\,119    & 05 Dec 2001 & 12 hr      & DA\,193              & 3C\,84 3C\,138, DA\,193 \\
 3C\,318    & 29 Sept 2001& 12 hr      & OQ\,208              & 3C\,279, 3C\,380 \\
 3C\,343    & 25 Oct 2001 & 12 hr      & OQ\,208              & 3C\,345, 3C\,380 \\
\hline
IFs       & MHz  & Bandwidth & Array     &                    &  \\
\hline  
IF1       & 4619 & 16\,MHz   & VLBA+VLA1 &                    &  \\      
IF2       & 4657 & 16\,MHz   & VLBA      &                    &  \\
IF3       & 4854 & 16\,MHz   & VLBA+VLA1 &                    &  \\
IF4       & 5094 & 16\,MHz   & VLBA      &                    &  \\ 
          &      &           &           &                    &  \\
IF1--4    & 8421 & 64\,MHz  & VLBA+VLA1 &                    &  \\
\hline
\hline
\end{tabular}
\end{table*}

The data were correlated with the National Radio Astronomy Observatory (NRAO) 
VLBA processor at Socorro and
calibrated, imaged and analysed using the AIPS package.  The flux
density and polarisation calibrations were done following the procedure
described in \citet{Rossetti09} for the source 3C\,147 observed using the
same system setup.  The flux density calibration
uncertainty is $\approx 3\%$.  The compact polarised sources
\object{DA\,193}, \object{3C\,345}, and \object{3C\,380} were used to
determine the instrumental polarisation (``D-term'') using the AIPS
task PCAL. The solution showed that the instrumental polarisation was
typically of order $1\%$.

\section{Results}
\label{sec:sources}
\subsection{3C\,119}

At different times, the optical identification for the radio source
3C\,119 has been called both a galaxy and a quasar, as noted by Fanti
et al. (1990) who classified it as a quasar.  Its light is
dominated by its nucleus, a morphology more typical of CSS quasars
\citep{devries97}. It has a reasonably  broad H$\beta$ profile and we
presently classify it as a possible quasar.  It has $m_{v}=20$ and
$z=1.023$ \citep{Eracleous94}, so that 1 mas corresponds to 8.086 pc.
A MERLIN  polarimetric image of 3C\,119 at 5\,GHz was obtained by
\citet{Ludke98} who studied a sample of CSS sources.  At their
resolution, it appears barely resolved. L\"{u}dke et al.  pointed out
that it shows extremely rapid depolarisation between 8.4 and 5\,GHz.
This was recently confirmed by \citet{Mantovani09} who detected
polarised emission of 8.8 and 5.9 \% at 10.45 and 8.35\,GHz
respectively, whereas at 4.85\,GHz the polarisation was below the
detection limit ($\sim$110\,mJy, corresponding to about 2.7\%) of their
Effelsberg 100-m telescope observations.

The first VLBI images of 3C\,119 with resolutions $\leq 10$\,mas were
made by \citet{Fanti86} at 18 and 6\,cm. They found at least four
components embedded in a complex, spiral, filamentary structure
suggesting that the low brightness emission was distributed in
filaments surrounding the brighter components. Global VLBI observations
at 18\,cm were also made by Nan et al. (1991a).  These observations,
combined with MERLIN data taken at the same time, confirmed the
existence of the four components found by \citet{Fanti86}.  These  were
labelled  A, B, C, and D, by Nan et al. (1991a) and we adopt the
same nomenclature here.  A further three, extended components of low
surface brightness, E, F and G, were also found by \citet{Nan91a}.
Together, these components account for 90\% of the total flux density
of the source.

VLBI polarimetry of 3C\,119 was first made by \citet{Nan99} using
the VLBA at three widely-separated frequencies in the 8.4\,GHz band.
The brightest jet component they found, component C, shows a smooth
rotation measure gradient of 2300 rad\,m$^{-2}$\,mas$^{-1}$ which
suggests a collision between the VLBI jet and a dense interstellar
cloud.

3C\,119 is included in the  MOJAVE \citep [Monitoring Of Jets in
Active galactic nuclei with VLBA Experiments;] [ ] {Lister09} monitoring
project at 15.4\,GHz. Four images from  2002, 2006, 2008 and 2009
are available in their data archive. At this frequency, components
A, B, and C are detected while the low brightness emission detected
by Nan et al. (1991a), is completely resolved out.

Our total intensity image of 3C\,119 at 4.8\,GHz made using all
four C-band IFs, and the image at 8.4\,GHz, are presented in
Fig.~\ref{fig:3c119-4IFs}.  The source exhibits a core-jet structure
with four of the seven components, (A--D), being detected.  The
three brightest components are not aligned, but lie along a jet
showing ``wiggles''. Components B and C are surrounded by a ``halo''
of spots, reminiscent of the low brightness emission in which the
jet itself is embedded.

Component A is almost point-like in the available VLBI images.  Using
flux density estimates for it made at almost the same epoch for
frequencies from 1.6 to 15.4\,GHz, we find a spectrum (see
Fig.~\ref{fig:3c119-spix}) typical of a Giga-Peaked Spectrum (GPS)
source, with a peak flux density around 5\,GHz.  In
Fig.~\ref{fig:3c119-spix}, the flux density measurements at frequencies
$>$ 4.5\,GHz were taken over a narrow range of dates.  The measurement
at 1.6\,GHz \citep{Nan91a} was adopted assuming negligible temporal
variability at that frequency. Component A is not detected at 50 cm
\citep{Nan91b}, with an upper limit of $\sim$40 mJy, consistent with
its inverted spectrum at low frequencies.  We note that the flux
density at 8.4\,GHz for this component dropped from 117 to 70\,mJy
between December 1994 \citep{Nan99} and December 2001 (present work).
Moreover, considering the observations made by MOJAVE at 15.4\,GHz, the
flux densities rose from 45\,mJy in October 2002 to 116\,mJy in March
2006, 149\,mJy in May 2008, and 173\,mJy in July 2009. Clearly, we are
dealing here with a variable GPS source.  Variability in a small
fraction of GPS sources has been pointed out by \citet{Jauncey03}.
Therefore, component A is almost certainly the core of 3C\,119.

%
\begin{figure*}[t]
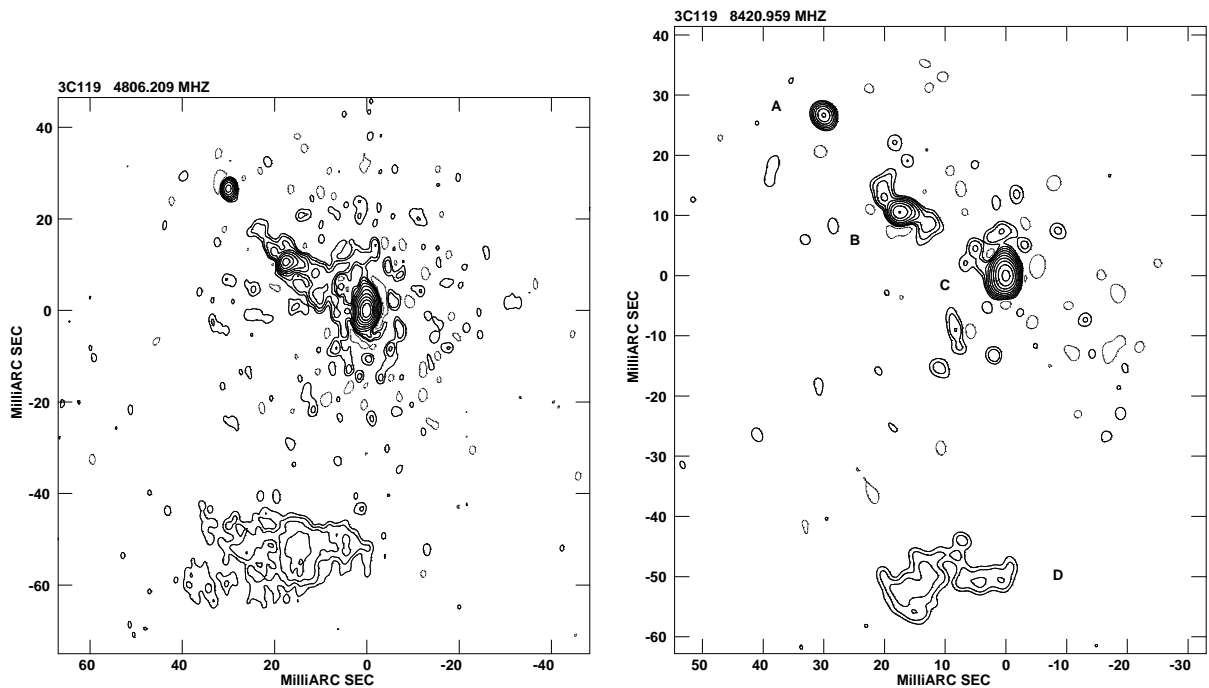

\addtocounter{figure}{+0}
\centering
\includegraphics[width=8cm]{3C119-ICL001-8.ps}
\includegraphics[width=8cm]{3C119-X-ICL-6-NOP.ps}
\caption{({\it Left}) The total intensity 
image of 3C\,119 at 4.8\,GHz  made by combining all four 
 C-band IFs, and ({\it right}) the total intensity 8.4\,GHz image  of 
 \object3C\,119.  The  restoring beams are 
$2.25\times 1.57$~mas$^{2}$ at $-11.2^{\degr}$,
 and $2.04\times 1.83$~mas$^{2}$ at $26.7^{\degr}$ respectively. 
The contour levels
 increase by factors of two from 1~mJy/beam for both images. 
Component designations are annotated on the 8.4\,GHz image.
\label{fig:3c119-4IFs}}
\end{figure*}

\vspace{0.3cm}
\begin{figure*}[t]
\addtocounter{figure}{+0}
\centering
\vspace{0.3cm}
\includegraphics[width=8cm,angle=-90]{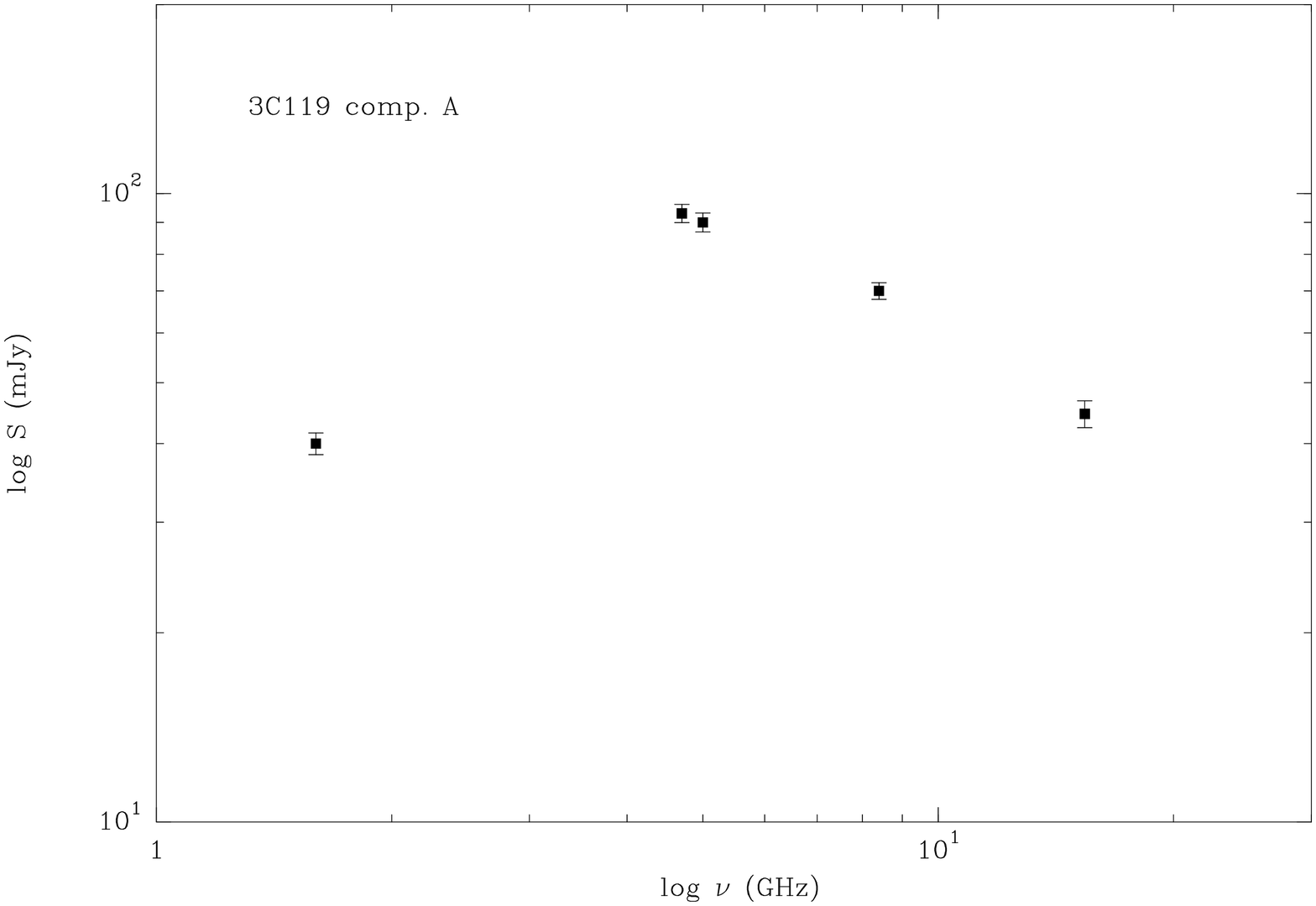}
\caption{Spectral index plot for component A made using flux density
measurements taken almost at the same epoch for frequencies $>$ 4.5\,GHz.
The measurement at 1.6\,GHz (Nan et al. 1991a) was adopted assuming 
negligible flux density variability at that frequency.   
\label{fig:3c119-spix}}
\end{figure*}
%

\citet{Mantovani09} found total flux densities for 3C\,119 of 4722
and 2677\,mJy at  4.8 and 8.4\,GHz respectively from their
Effelsberg 100-m observations.  The total flux density in our
4.8\,GHz combined-IF image is  2838\,mJy. At 8.4\,GHz we
obtain 1531\,mJy. The peak flux density is located in component C,
being 987 and 660\,mJy at 4.8 and 8.4\,GHz respectively.

\begin{figure*}[t]
\addtocounter{figure}{+0}
\centering
\includegraphics[width=8cm]{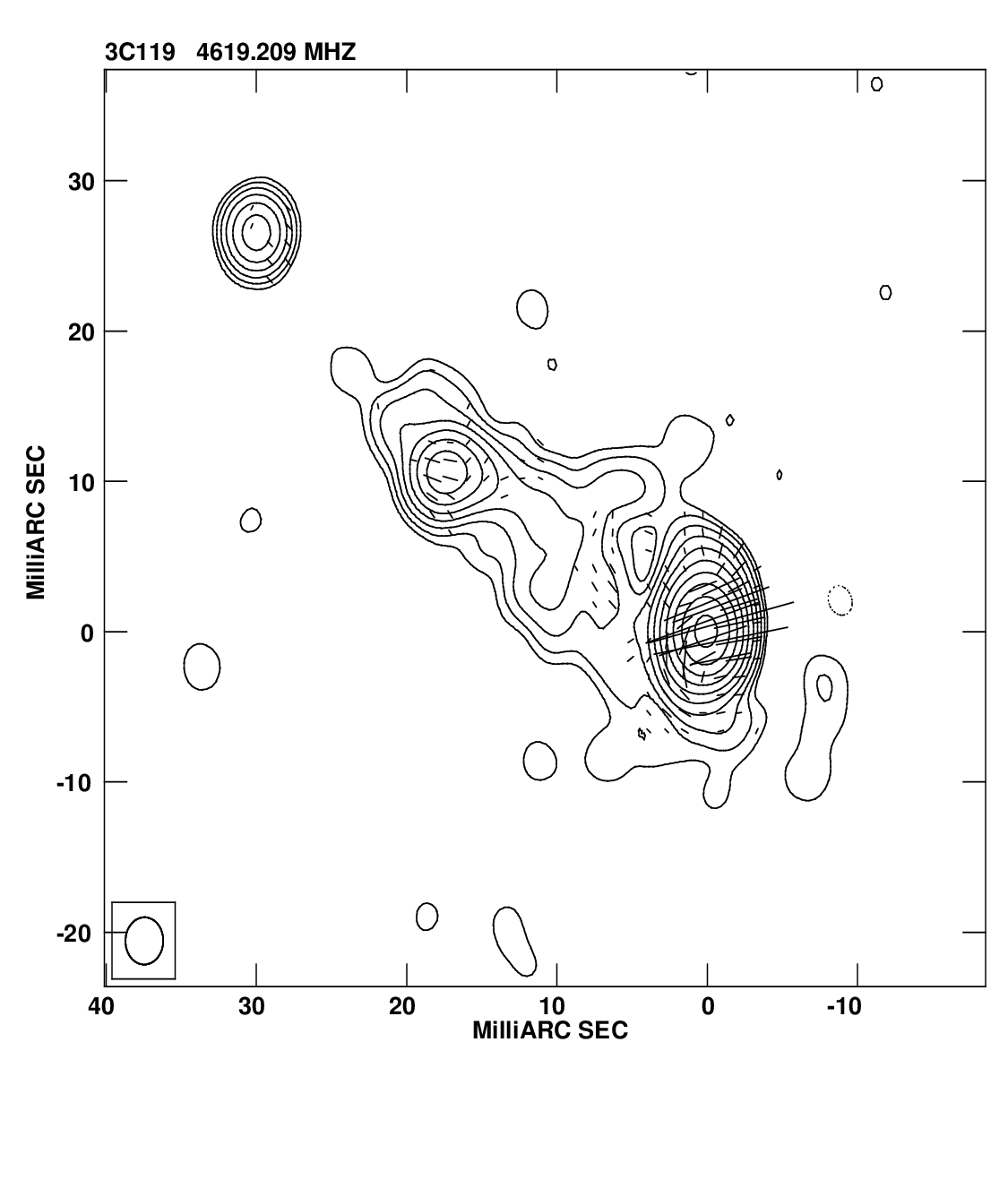}
\includegraphics[width=8cm]{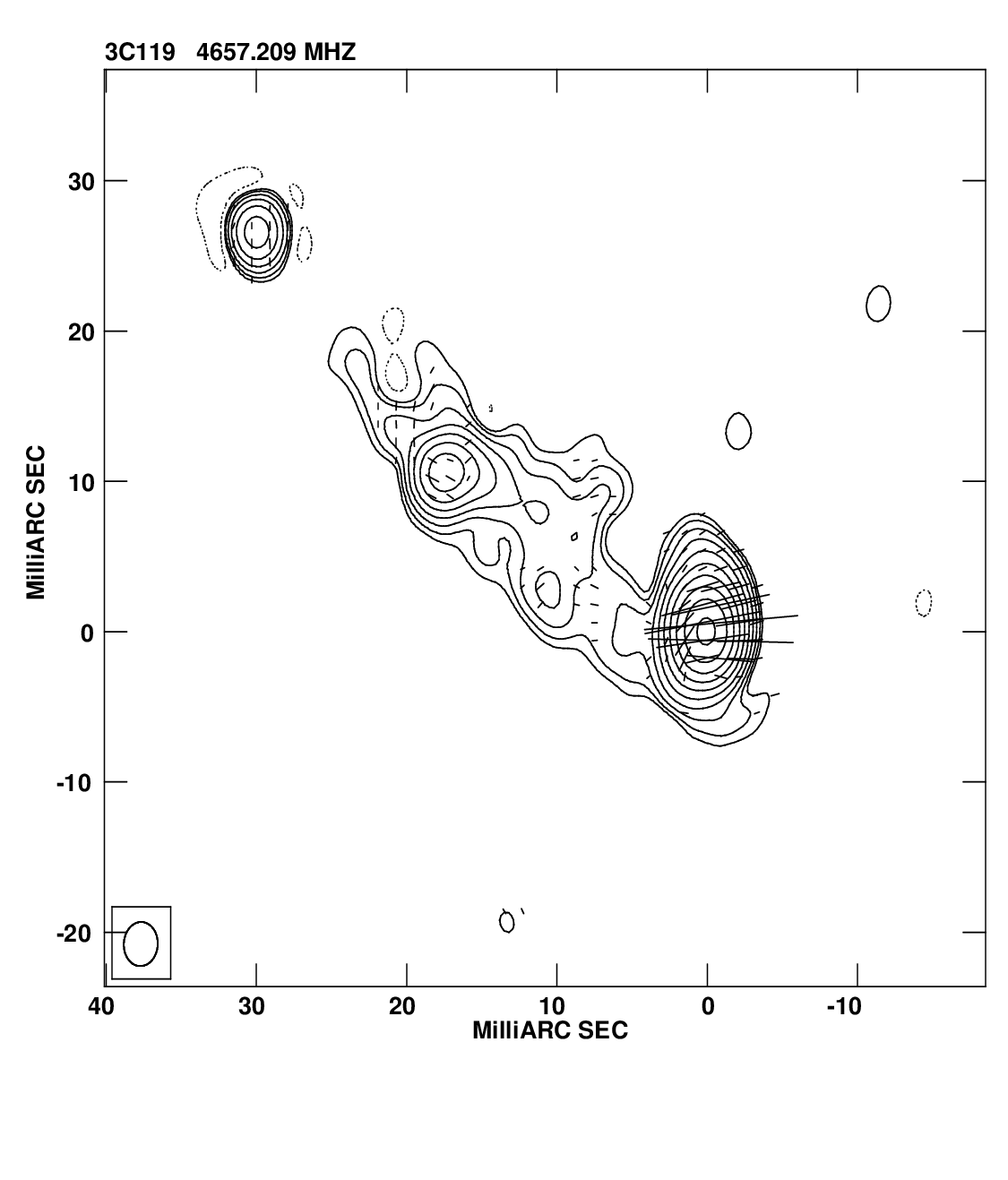}
\includegraphics[width=8cm]{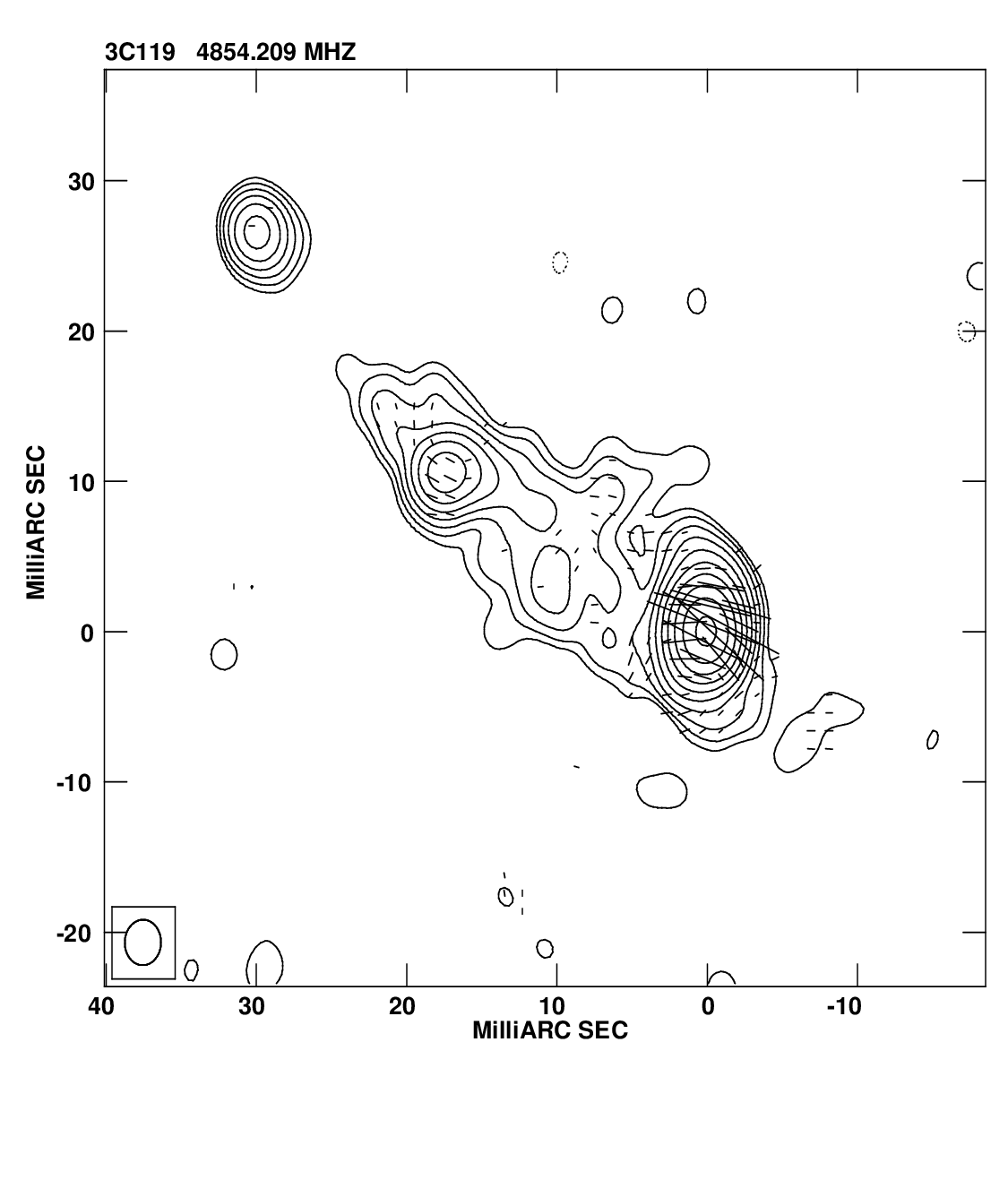}
\includegraphics[width=8cm]{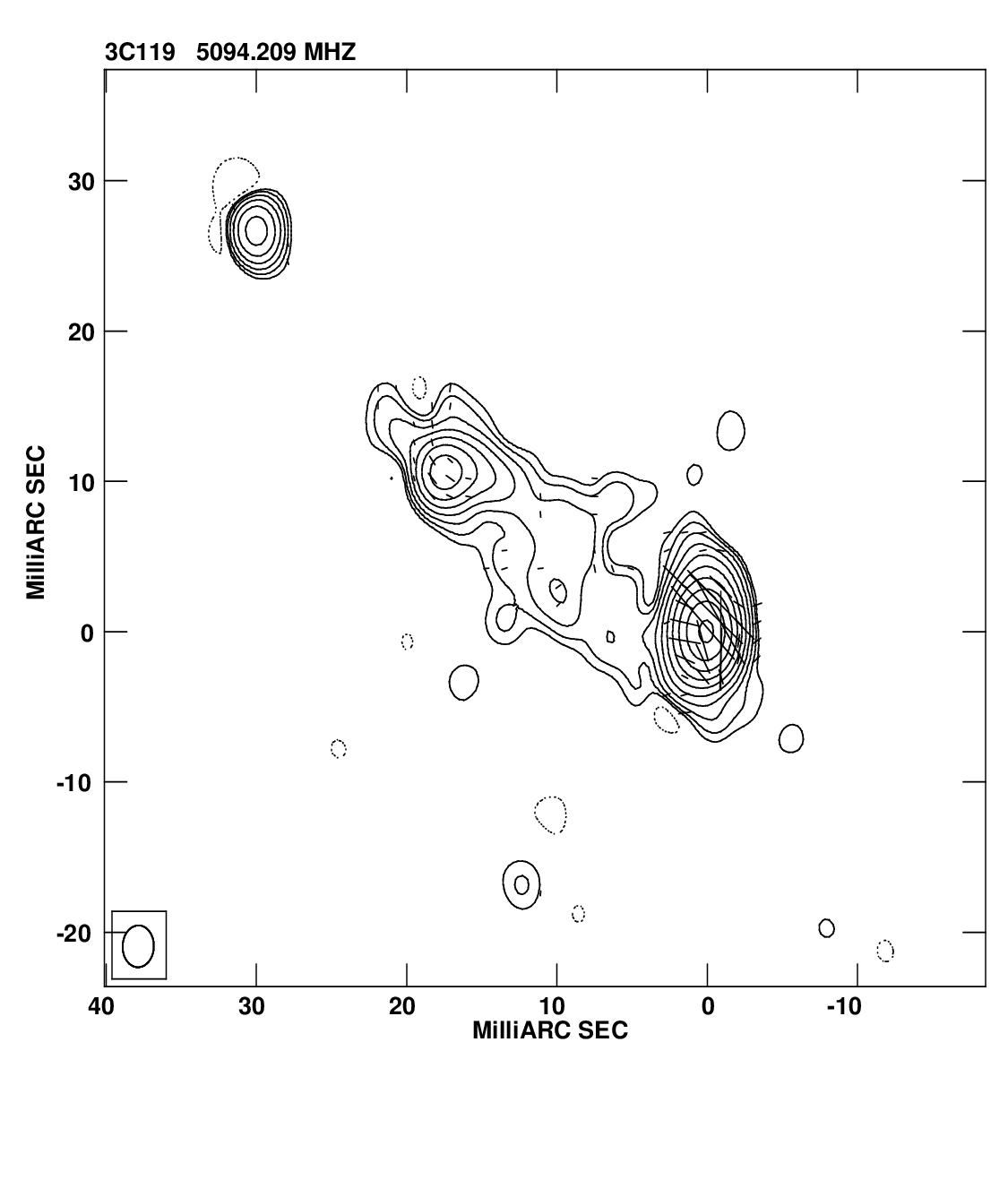}
\caption{The total intensity contours for 3C\,119\,A-C 
for the four C-band IFs with the polarisation E vectors
superimposed. Contour levels increase by factors of two from 2~mJy/beam.
A vector of 1\,mas corresponds to 1.67\,mJy/beam of polarised emission.
The convolution beam is 3$\times$2 mas$^2$ at PA 0$^{\degr}$.
\label{fig:3c119-polC}}
\end{figure*}
\begin{figure*}[t]
\addtocounter{figure}{+0}
\centering
\includegraphics[width=8cm]{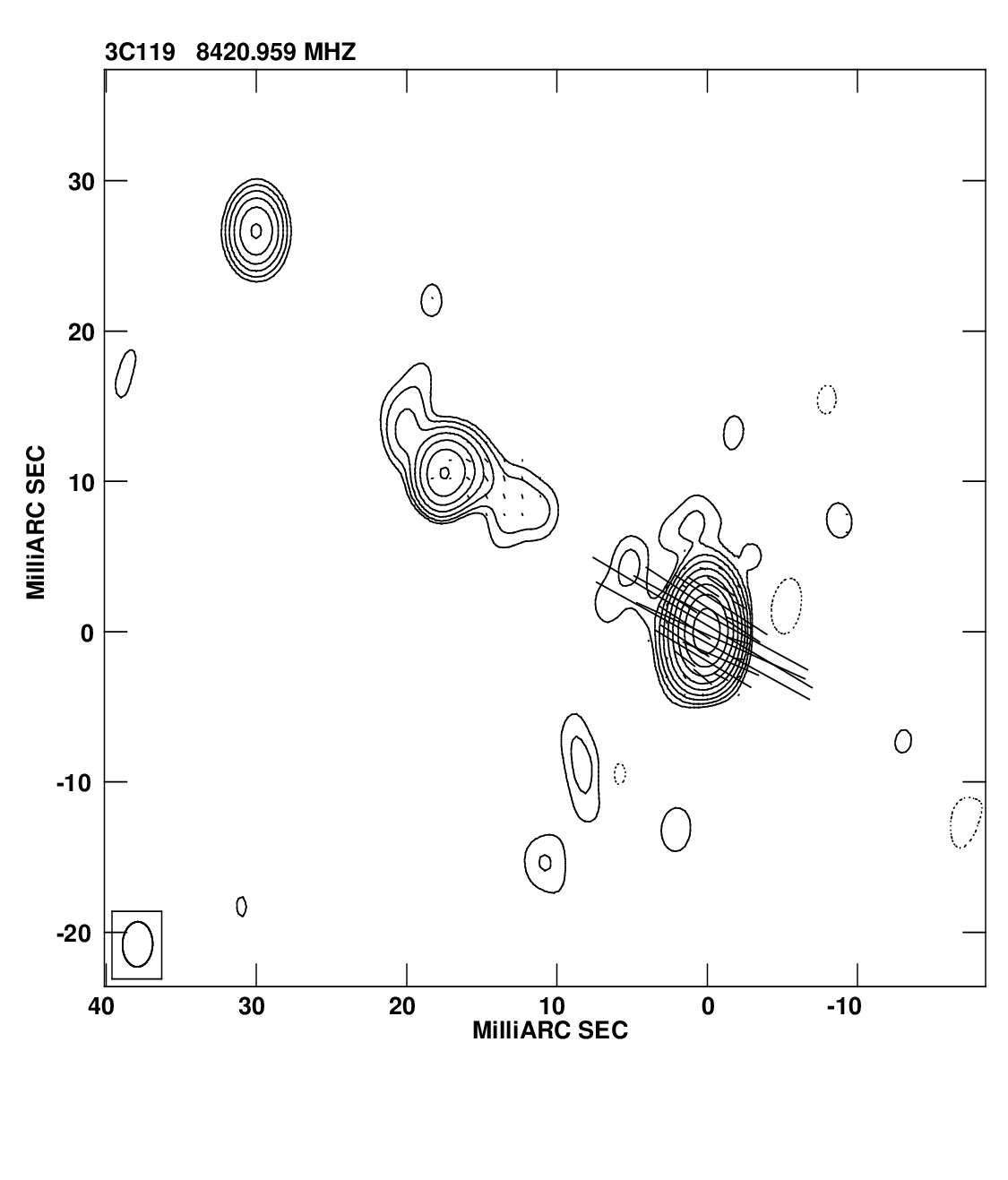}
\caption{The total intensity contours for 3C\,119\,A-C 
at X-band with the polarisation E vectors
superimposed. Contour levels increase by factors of two from 2~mJy/beam.
A vector of 1\,mas corresponds to 6.7\,mJy/beam of 
polarised emission. The convolution beam is 3$\times$2 mas$^2$ at PA 0$^{\degr}$.
\label{fig:3c119-polX}}
\end{figure*}

\begin{table*}[htbp]
\centering
 \caption{Polarimetric parameters for the  inner jet of 3C\,119.}
 \label{tab:fluxes}
\tabcolsep 1.4mm
 \begin{tabular}{ccccccccc}
 \hline
 \hline
Comp. A & &          &           &              &          &               &       &\\
\hline 
IFs    &      & S$_{tot}$ & S$_{peak}$ &rms($1\sigma$)& S$_{pol}$& rms($1\sigma$)&  $m$  & EVPA  \\
       & MHz  & mJy      & mJy/b       & mJy/b        & mJy      & mJy/b         &  \%   & deg   \\
\hline
       &      &          &           &              &          &               &       &\\  
IF1    & 4619 &  93.7    &  93.2     &   0.7        & $<$0.6   &   0.11        &$<$0.6 &          \\      
IF2    & 4657 &  92.0    &  92.0     &   0.6        &$\sim$0.9 &   0.11        &$\sim$1& $\sim$0  \\
IF3    & 4854 &  95.0    &  89.4     &   0.6        & $<$0.6   &   0.12        &$<$0.6 &          \\
IF4    & 5094 &  90.5    &  90.5     &   0.6        & $<$0.7   &   0.13        &$<$0.7 &          \\
       &      &          &           &              &          &
 &   &   \\
IF1--4 & 8409 &  70.0    &  69.4     &   0.1        &          &   0.1         &       &       \\
       &      &          &           &               &         &               &       &     \\
\hline
Comp.B & &          &           &              &          &               &       &   \\
\hline
       &      &          &           &              &          &               &       &   \\  
IF1    & 4619 &  251.3   &  216.3    &   0.7        &  1.7     &   0.11        & 0.7   & 73$\pm2$      \\      
IF2    & 4657 &  267.2   &  207.1    &   0.6        &  1.4     &   0.11        & 0.5   & 62$\pm1$      \\
IF3    & 4854 &  272.0   &  211.2    &   0.6        &  1.8     &   0.12        & 0.7   & 59$\pm3$      \\
IF4    & 5094 &  273.8   &  199.2    &   0.6        &  1.6     &   0.13        & 0.6   & 44$\pm6$      \\
      & & & & &  &    & & \\
IF1--4 & 8409 &  155.6   &  134.2    &   0.1        &  0.8     &   0.09        & 0.5   & --77$\pm5$      \\
       &      &          &           &               &         &               &       &     \\
\hline
Comp.C & &          &           &              &          &               &       &\\
\hline 
       &      &          &           &               &         &               &       &     \\
IF1    & 4619 &  1831.4  &  1255.7   &   0.7         &  19.6   &   0.11        & 1.1   & 105$\pm2$      \\      
IF2    & 4657 &  1827.6  &  1204.3   &   0.6         &  24.2   &   0.11        & 1.3   & 92$\pm6$      \\
IF3    & 4854 &  1819.7  &  1229.1   &   0.6         &  24.3   &   0.12        & 1.3   & 62$\pm6$      \\
IF4    & 5094 &  1800.8  &  1157.2   &   0.6         &  22.1   &   0.13        & 1.2   & 40$\pm5$      \\
  & &   &  &  & & & & \\
IF1--4 & 8409 &  1121.4  &   806.2   &   0.1         & 161.1   &   0.09        &14.4   & 61$\pm2$      \\
       &      &          &           &               &         &               &       &     \\
\hline
\hline
\end{tabular}
\normalfont
\smallskip\noindent
\flushleft{\normalsize {
The values for the three brightest components along the core-jet structure
are organised as follows: 
column 1, IF number; column 2, observing frequency; 
column 3, total flux density; column 4, peak flux density; 
column 5, 1$\sigma$ rms noise;
column 6, polarised flux density; column 7, 1$\sigma$ rms noise;
column 8, percentage polarisation; column 9, electric vector
position angle.}}
\end{table*}

\subsubsection{Polarised emission from the inner jet of 3C\,119}
In order to compare the 5 and 8.4~GHz results, the images of 3C\,119
have been convolved to a resolution of 3.0 $\times$ 2.0\,mas$^{2}$ at
0$^{\degr}$.  We detect polarised emission along the jet structure.
However, components A and D show no, or at best marginal, polarised
emission above our detection limits at both 4.8 or 8.4\,GHz.  At
C-band we have imaged the jet for each of the four individual
IFs.

Fig.~\ref{fig:3c119-polC} shows the total-intensity and polarisation
structure of the jet at C-band for the four individual IFs, while
Fig.~\ref{fig:3c119-polX} shows the jet structure at X-band.  The
parameters derived from these images are listed in
Table~\ref{tab:fluxes}.  Flux densities were determined using the AIPS
task IMEAN on the same region of the P and I images. In
Table~\ref{tab:fluxes}, the total and peak flux densities, S, the
rms noises, the polarised integrated flux densities, S$_{pol}$ and its
rms noise, the percentage polarisation, $m$, and the electric vector
position angle (EVPA), $\chi$, are listed for the four C-band IFs and
X-band. For the core component A, which appears unresolved, we have
listed the values at the pixel of maximum total intensity.

Plots of $\chi$ against wavelength squared for components B and
C are shown in Fig.~\ref{fig:3c119-RMplot}. These were derived as
follows.  The median values and associated errors were computed for a
box of five-by-five pixels around the peaks of polarised emission.
At 15.4\,GHz the EVPA for both components was derived from the image
made available by the MOJAVE archive from observations taken in 2002.
These values have an accuracy of better than 5$^{\degr}$ 
\citep{Lister05}. 

Both components show high
values for the rotation measure. We derive {\it RM} = 884
rad\,m$^{-2}$ (3618 rad\,m$^{-2}$ in the source rest frame) for
component B, whereas component C yields {\it RM} = 1373 rad\,m$^{-2}$
(5620 rad\,m$^{-2}$ in the source rest frame).  We note that the
EVPAs reported by \citet{Nan99} for component C at the three
sub-bands of their VLBA X-band observations are well aligned with the
linear fit in Fig.~\ref{fig:3c119-RMplot}.  They did not detect
polarised emission in the region close to the peak of emission for
component B, in contrast to our more sensitive observations. Polarised
emission was detected by both sets of observations for the region
immediately to the south-west of compact component B, where almost the
same values were found for the EVPAs.

\vspace{0.3cm}
\begin{figure*}[t]
\addtocounter{figure}{+0}
\centering
\vspace{0.3cm}
\includegraphics[width=8cm,angle=-90]{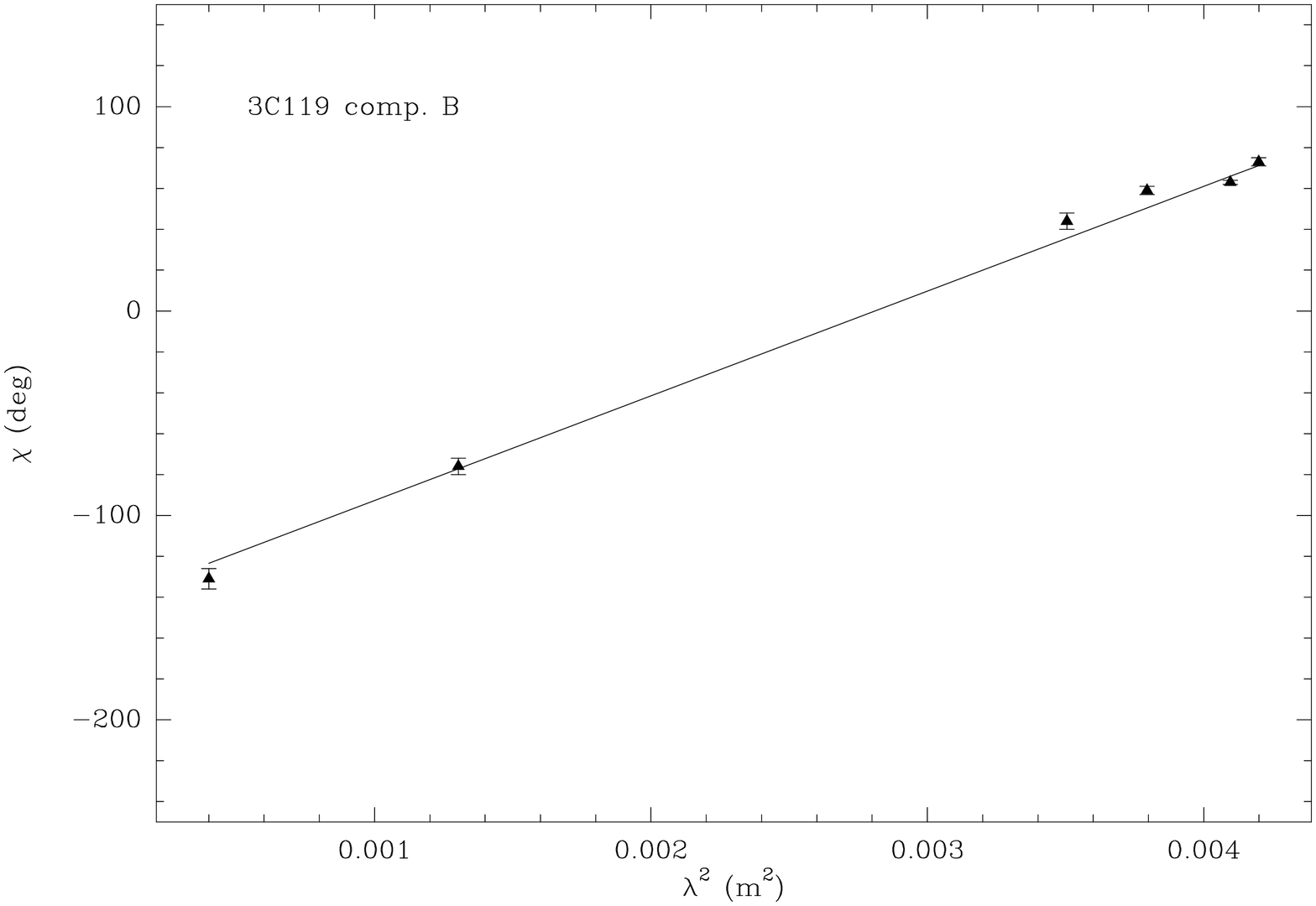}
\includegraphics[width=8cm,angle=-90]{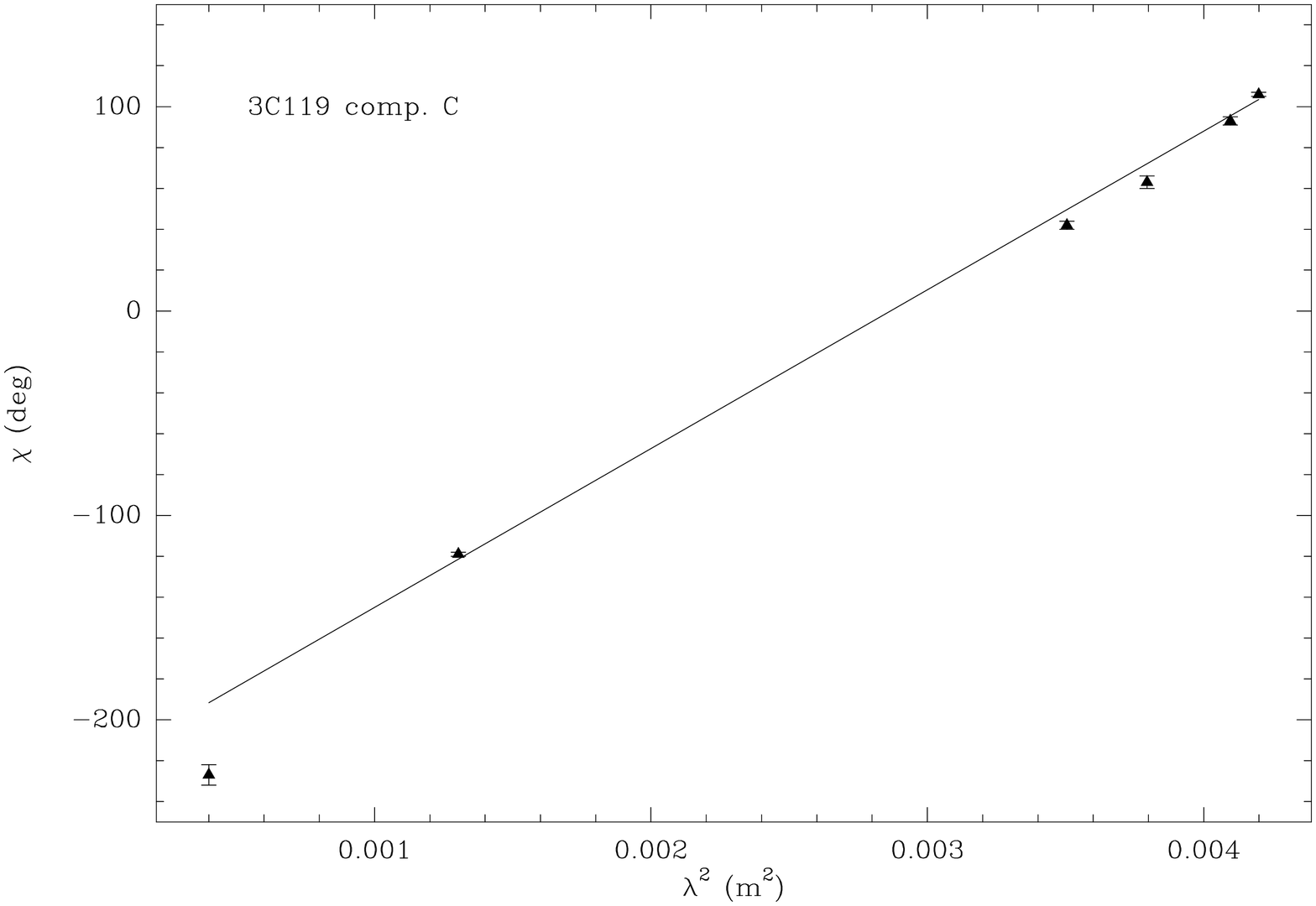}
\vspace{0.3cm}
\caption{Plots of the observed $\chi$ values for components B ({\it top}) and
C ({\it bottom}) of 3C\,119 as a function of $\lambda^2$ for the six available 
frequencies.
\label{fig:3c119-RMplot}}
\end{figure*}
\vspace{0.3cm}
\begin{figure*}[t]
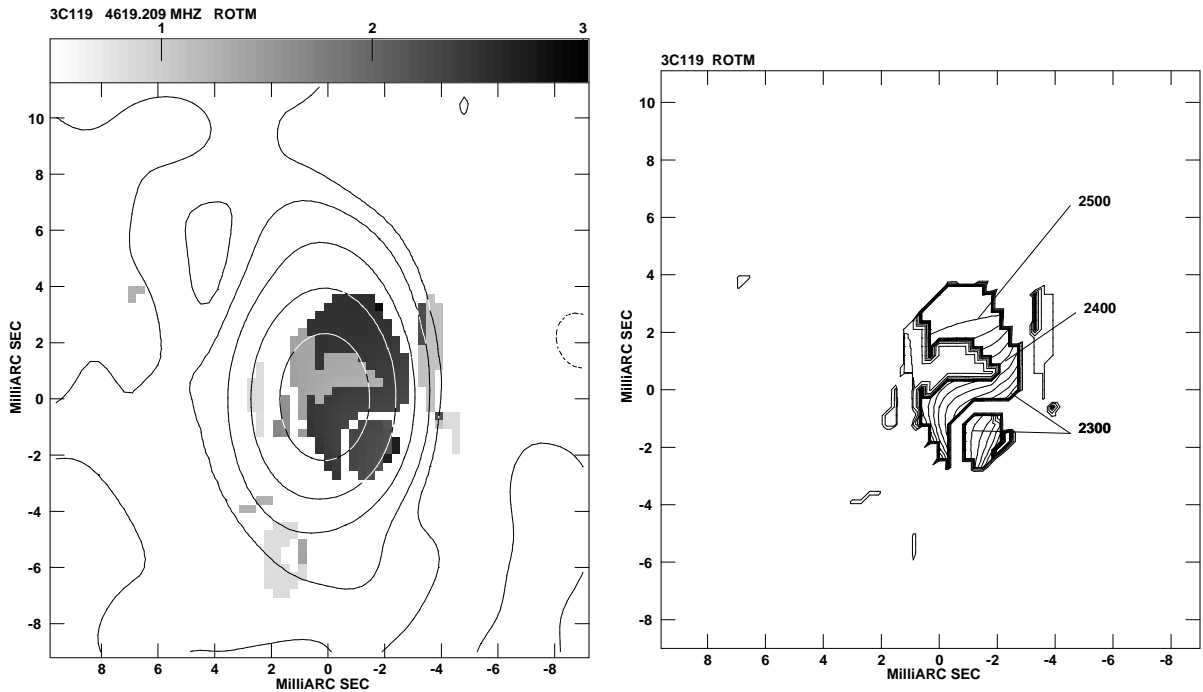

\addtocounter{figure}{+0}
\centering
\vspace{0.3cm}
\includegraphics[width=8cm]{3C119-RM-SMA.PS}
\includegraphics[width=8cm]{3C119-RM-SMA-PL.PS}
\vspace{0.3cm}
\caption{The derived distribution of {\it RM} for 3C\,119~C 
as a grey-scale plot 
({\it left}) and as a contour plot ({\it right}).
The range of {\it RM} in the left panel is from 500 to 3000 rad m$^{-2}$,
and the distribution is overlaid on the continuum image at 4619.2 MHz. 
The continuum contour levels 
are $-$2, 2, 8, 32, 128 and 512 mJy beam$^{-1}$. On the right panel the {\it RM} 
contour levels are
100$\times$(10, 14, 18, 20, 22, 22.25, 22.50, 22.75, 23, 23.25, 23.50, 23.75, 
24, 24.25, 24.50, 24.75, 25) rad m$^{-2}$.
\label{fig:3c119-RMdistr}}
\end{figure*}

We have also plotted the distribution of {\it RM} around the emission
peak in component C (Fig.~\ref{fig:3c119-RMdistr}).  These plots
were produced using the AIPS task RM which requires EVPAs for up
to four frequencies to compute the {\it RM} values.  In this case we have
selected frequencies of 4619, 4854, 5094, and 8421\,MHz for which we
have images available.  Fig.~\ref{fig:3c119-RMdistr} shows that
in a small area close to the peak of polarised emission there are
values about 1400 rad\,m$^{-2}$ (see Fig.~\ref{fig:3c119-RMplot}). This
area is surrounded by a region with much higher {\it RM}s lying in the
range 2300 to 2500 rad\,m$^{-2}$ (9400-10200 rad\,m$^{-2}$ in the
source rest frame).  This region coincides with a change in the
direction of the inner jet of 3C\,119.  The value of {\it RM} reported for
the integrated emission of 3C\,119 by \citet{Mantovani09} is 
1928 rad\,m$^{-2}$. Clearly,  C is the dominant component,
strongly influencing the polarimetric parameters of the source.

Fig.~\ref{fig:3c119-RMdistr}~(left) shows in grey-scale the {\it RM}
distribution for \object{3C\,119} derived from the four frequencies (4619, 
4854, 5094, and 8421~MHz) overlaid on the total-intensity contour
image, while the panel on the right shows the {\it RM} contours.  No 
redshift corrections have been applied, so the {\it RM}s in the
rest frame of the source will be higher by a factor of $(1+z)^2\approx 4$.

\subsection{3C\,318}

3C\,318 is a radio galaxy at a red-shift of 1.574 (1 mas = 8.554 pc).
Single-dish observations of the object were recently made by
\citet{Mantovani09} who measured flux densities of 764\,mJy at
4.8\,GHz and  417\,mJy at 8.4\,GHz.  The source is 3.4\% and 6.6\%
polarised at 4.8  and 8.4\,GHz respectively. At lower frequencies
the polarisation drops below the detection limit of these Effelsberg
observations. The {\it RM} derived by Mantovani et al. is 498
rad\,m$^{-2}$.  Using the measurements of \citet{Tabara80} yields an
{\it RM} of 342 rad\,m$^{-2}$.  A similar value for the brightest
component (420 rad\,m$^{-2}$) is reported by \citet{Taylor92} from VLA
observations with angular resolution of 0.4$\arcsec$.  Taylor et al. found a 
polarisation percentage of $\approx$10\% at 8.4\,GHz.

A higher resolution L-band image of 3C\,318 was made by \citet {Spencer91}
using combined MERLIN and EVN observations at a resolution of
35$\times$33 mas$^2$. The component B detected by \citet{Akujor91} with
MERLIN at 5\,GHz was resolved out, while the two northern components
show considerable structure and bright peaks.  The total flux density
detected by \citet{Akujor91} was 543\,mJy.  \citet{A-G95} made
polarimetric observations of 3C\,318 at 8.4\,GHz with the VLA, and
detected polarised emission ($\approx$10\%) mainly from the two
northern components. VLA observations were also made by
\citet{Breugel92} at 15  and 22\,GHz. They found the northern
region, in which the two component are merged, to be 17\%
polarised. Further polarimetric observations were made by
\citet{Ludke98} with MERLIN who found the brightest component to be
3.6\% polarised at 5\,GHz. In this image, 3C\,318 shows a core-jet
structure on one side and a lobe on the other. It is suggested that
component K2 \citep {Spencer91} might be the core, although it
appears resolved and does not have a flat spectrum.

3C\,318 appears to be heavily resolved by our 4.8\,GHz VLBA
observations.  Fringes were not detected at 8.4\,GHz. Our 5\,GHz image
was made using the full bandwidth and, in order to compare with
previous VLBI observations, with a convolution beam of 35$\times$33
mas$^2$ in PA 46$^{\circ}$ \citep{Spencer91}.  3C\,318 appears
elongated in roughly the North-South direction, with the southern side
showing a ``wiggling'' structure (Fig.~\ref{fig:3c318-C}). We follow
the nomenclature of Spencer et al.  (1991) who found four components
named K1, K2, K3, and A.  About 61\% of the flux density found by
\citet{Ludke98} is recovered here.  The source parameters derived from
our image are summarised in Table~\ref{tab:3c318-par}, the flux density
values having been obtained using the AIPS task TVSTAT.

\vspace{0.3cm}
\begin{figure*}[t]
\addtocounter{figure}{+0}
\centering
\vspace{0.3cm}
\includegraphics[width=8cm]{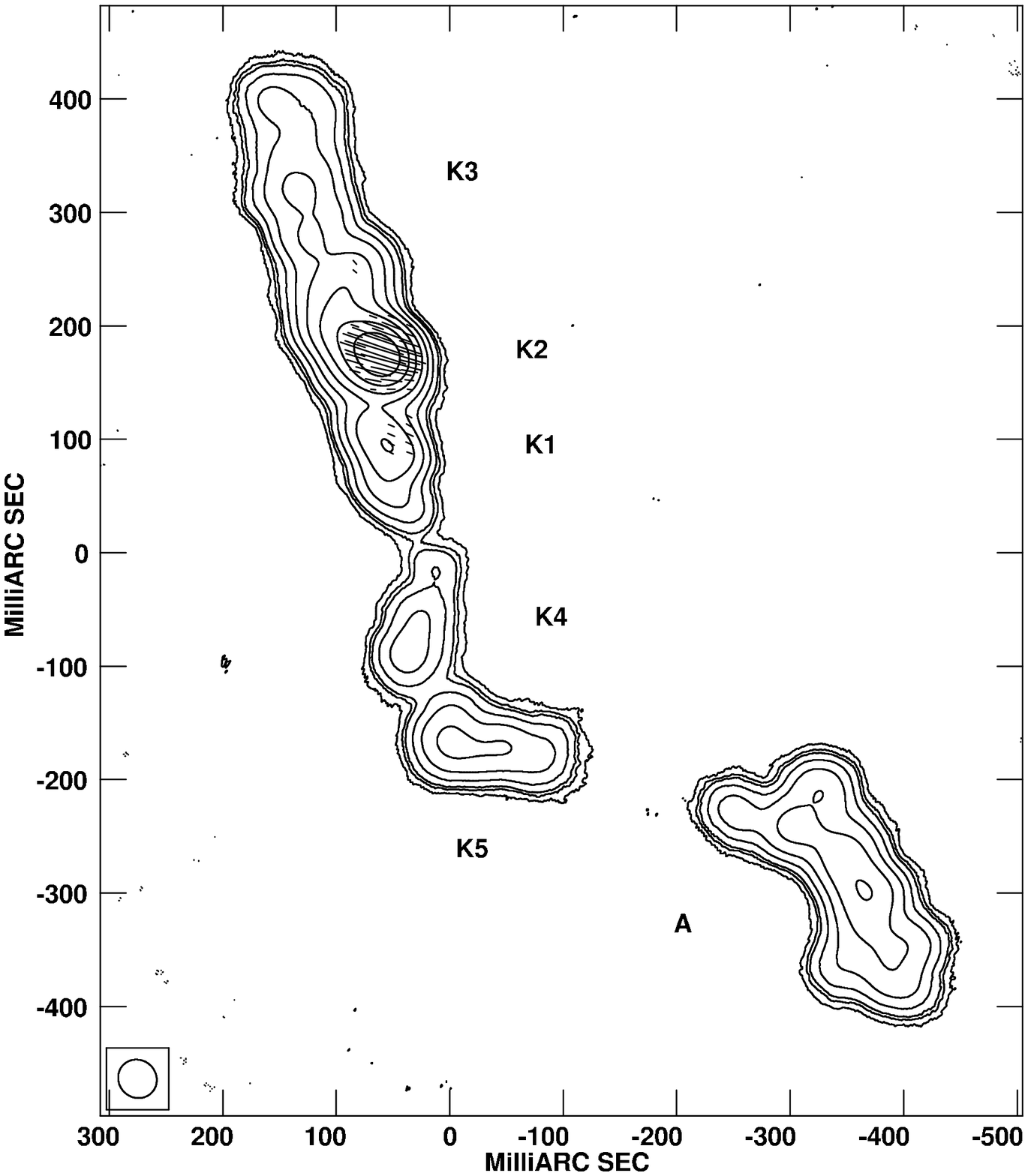}
\vspace{0.3cm}
\caption{The VLBA image of 3C\,318 at 4.8\,GHz. Contour levels are --0.35,
0.35, 0.7, 1, 2, 4, 8, 16, 32, 64, 128~mJy/beam.
A vector of 1\,mas corresponds to 0.1\,mJy/beam of polarised emission.
The convolution beam is 35$\times$33 mas$^2$ at PA 46$^{\degr}$.
The component designations are annotated on the image.
\label{fig:3c318-C}}
\end{figure*}   

In our image, we find two other extended components, which we
designate K4 and K5.  These were not detected by the less
sensitive observations of \citet{Spencer91}.

\begin{table*}[htbp]
\centering
 \caption{Parameters for 3C\,318 at 4.8\,GHz with a
resolution of $35 \times 33$~mas$^{2}$ at 46$^{\degr}$. 
The rms noise ($1\sigma$) is 0.1 and 0.08 mJy/b for total and polarised 
flux density measurements respectively.
 \label{tab:3c318-par}}
\tabcolsep 1.4mm
 \begin{tabular}{cccccc}
 \hline
 \hline
Comp.    & S$_{tot}$ & S$_{peak}$ & S$_{pol}$  &  $m$  & EVPA       \\
         & mJy       & mJy        & mJy        &  \%   & deg        \\
\hline
         &          &            &             &       &            \\  
K1       &     35.2 &  16.4      &   0.8       &  2.2  &  72$\pm$5   \\
K2       &    228.4 & 122.2      &   8.0       &  3.5  &  69$\pm$5   \\
K3       &     47.9 &  9.8       &             &       &             \\
K3       &     47.9 &  9.8       &             &       &              \\
K4       &     15.0 &  5.9       &             &       &              \\
K5       &     34.7 &  9.9       &             &       &              \\
A        &    107.5 & 16.8       &             &       &              \\
\hline
\hline
\end{tabular}
\end{table*}

The brightest component, K2, is responsible for almost all of the
polarised emission from 3C\,318, and the magnetic field orientation appears
almost constant over the region. A similar behaviour, but with
different position angles, is presented by all available
polarimetric images.  Thus, we can derive the values of the rotation
measure from these interferometric observations made at different
frequencies. Values of the EVPA are taken from \citet{Ludke98} at
4996\,MHz, Akujor \& Garrington (1995) at 8414\,MHz, \citet{Taylor92} at
8515\,MHz, and \citet{Breugel92} at 15\,GHz, as well as from the
present work, and are plotted in Fig.~\ref{fig:3c318-rm-plot}.  
EVPA accuracies for these observations were not given, and values of 
$\pm2^{\degr}$ have been assumed for both MERLIN and VLA EVPA determinations.
The {\it RM} computed is $\approx$457 rad\,m$^{-2}$, not far from
the values of 420 rad\,m$^{-2}$ derived by \citet{Taylor92}, and
498 rad\,m$^{-2}$ derived by \citet{Mantovani09}. Again we are
dealing with a CSS showing a very high {\it RM} of $\approx$3030
rad\,m$^{-2}$ in the source rest frame at sub-arcsecond resolution.

%
\begin{figure*}[t]
\addtocounter{figure}{+0}
\centering
\vspace{0.3cm}
\includegraphics[width=8cm,angle=-90]{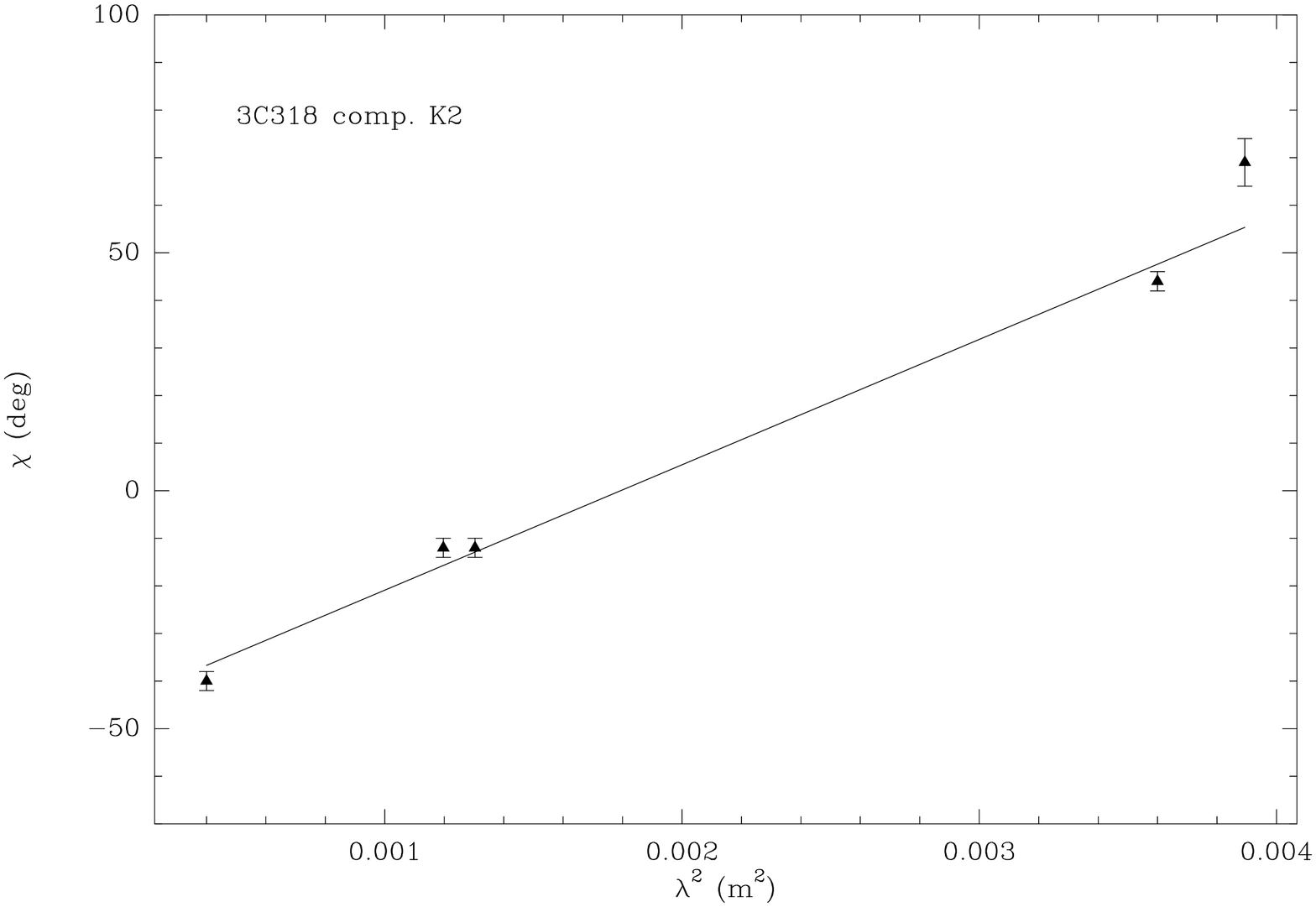}
\vspace{0.3cm}
\caption{A plot of the observed $\chi$ values for component K2 of 3C\,318 
as a function of $\lambda^2$ for the five available frequencies.
\label{fig:3c318-rm-plot}}
\end{figure*}   

\subsection{3C\,343}   

3C\,343 is associated with a quasar at $z$=0.988 (1 mas = 8.017 pc). From their
single dish observations, \citet{Mantovani09} reported the detection
of polarised flux density for this source only at 10.45\,GHz at a level
of 1.4\%. A low level of polarised emission (0.8\%) was also reported
by \citet{Ludke98} from their MERLIN observations at 5\,GHz in which
the source appears barely resolved with a total flux density of
1434\,mJy. Polarised emission is detected only in its eastern
extension.  Previous VLBI observations were made by \citet{Fanti85} and
\citet{Nan91b} at 1.6\,GHz and 608\,MHz respectively. The mas structure
of the source is quite peculiar, resembling a `fried egg'. 

Our total-intensity VLBA images of 3C\,343 for the different C-band
IFs are shown in Fig.~\ref{fig:3c343-Cband}, while the VLBA image at
8.4\,GHz is shown in Fig.~\ref{fig:3c343-Xband}.  The wide-band 5\,GHz
VLBA system allowed us to use four 8\,MHz IF channels well separated in
frequency.  Only C-band IFs 1 and 3 contain data from the VLA antenna
which was included in the observing array.  Two bright, compact
components, A and B, are surrounded by weak, diffuse emission.  The two
images made from data including the VLA antenna show the diffuse
emission, while the two images lacking baselines to the VLA antenna
show the two prominent components, A and B, and the ridge of emission
which resembles a curved radio jet.  Both A and B have steep spectral
indices between 4.8 and 8.4\,GHz.  The flux densities for these
components have been obtained with the AIPS task JMFIT, and component A
(the eastern one) has $\alpha\approx$1.6 (1.3 using the peak flux
densities), while component B has $\alpha\approx$1.2 (0.8 using the
peak flux densities).  Polarised emission is found from both of
these components and along the emission region between them.  This
all suggests it to be unlikely that either A or B is the core of
3C\,343.

From the 8.4\,GHz image, we note that a new compact component is
clearly visible directly to the east of component A.  This feature,
which we label C in Fig.10, is well separated from the brighter
component A. Using the AIPS task JMFIT, we obtain total and peak flux
densities of 7.7 and 3.5\,mJy respectively for Component C.  It is not
seen on any of the C-band images of the same resolution, including that
made by combining all four IFs, which means a 3$\sigma$ upper limit peak flux 
density of 1\,mJy.  This indicates an inverted spectrum
for C between 4.8 and 8\,GHz, making it a candidate for being the core
of 3C\,343. Moreover, component C is situated nicely on the curved
track of the jet as seen in Figs. 9 b and d.

The parameters for components A and B are summarised in 
Table~\ref{tab:3c343-par}. The component peak and total flux densities,
along with their deconvolved angular dimensions, have been
obtained using the AIPS task JMFIT.  The degree of polarisation has
been estimated close to the peaks of emission in A and B. However,
there appears to be large changes in the EVPAs close to the peaks
of emission. Therefore, we have not quoted a single value in 
Table~\ref{tab:3c343-par} 
but  refer the reader to the distributions of EVPAs in 
Fig.~\ref{fig:3c343-Cband}. 

\begin{table*}[htbp]
\centering
 \caption{Polarimetric parameters for the components A and B of 3C\,343.
 \label{tab:3c343-par}}
\tabcolsep 1.4mm
 \begin{tabular}{cccccccccccr}
 \hline
 \hline
Comp. A & &          &           &              &          &               &       &\\
\hline 
IFs    &      & S$_{tot}$ & S$_{peak}$ &rms($1\sigma$)& S$_{pol}$& rms($1\sigma$)& maj.ax & min.ax. & PA &  $m$  & EVPA  \\
       & MHz  & mJy      & mJy/b       & mJy/b        & mJy      & mJy/b       &  mas   & mas     & deg&  \%   & deg   \\
\hline
       &      &          &           &              &          &                &      &      &          &       &           \\  
IF1    & 4619 &  102.1   &  30.3     &   0.4        &  3.7     &   0.2         
& 16.8 & 8.4  & 108      &   3.6 & see Fig.~\ref{fig:3c343-Cband}  \\  
IF2    & 4657 &  80.0    &  33.3     &   0.6        &  2.6     &   0.2         
& 12.5 & 6.5  & 112      &   3.2 &see Fig.~\ref{fig:3c343-Cband}  \\  
IF3    & 4854 &  80.3    &  33.5     &   0.3        &  4.8     &   0.2         
& 11.9 & 7.1  & 113      &   6.0 &see Fig.~\ref{fig:3c343-Cband}  \\  
IF4    & 5094 &  54.7    &  36.0     &   0.4        &  2.6     &   0.2         
& 6.5  & 5.0  & 131      &   4.7 &see Fig.~\ref{fig:3c343-Cband}  \\  
       &      &          &           &              &          &               
&      &      &          &       &                \\
IF1--4 & 8409 &  33.0    &  16.3     &   0.2        &  1.6     &   0.1         
& 8.7  & 7.5  & 105      &   4.8 & $-5\pm3$       \\
       &      &          &           &               &         &               
&      &      &          &       &     \\
\hline
Comp.B &      &          &           &              &          &               
&      &      &          &       &     \\
\hline
       &      &          &           &              &          &               
&      &      &          &       &     \\  
IF1    & 4619 &  470.5   &   77.2    &   0.4        &  2.7     &   0.2        
& 19.4 & 16.7 & 117      &   0.6 &see Fig.~\ref{fig:3c343-Cband}  \\  
IF2    & 4657 &  305.2   &   69.3    &   0.6        &  3.6     &   0.2        
& 16.6 & 13.1 & 128      &    1.2 &see Fig.~\ref{fig:3c343-Cband}  \\  
IF3    & 4854 &  391.9   &   80.6    &   0.3        &  4.3     &   0.2        
& 16.8 & 14.7 & 149      &    1.1 &see Fig.~\ref{fig:3c343-Cband}  \\  
IF4    & 5094 &  208.8   &   68.5    &   0.4        &  3.2     &   0.2        
& 12.5 & 10.4 & 131      &    1.5 &see Fig.~\ref{fig:3c343-Cband}  \\  
       &      &          &           &              &          &              
&      &      &          &           &             \\
IF1--4 & 8409 &  198.7   &   51.4    &   0.2        &  5.9     &   0.1        
& 15.8 & 11.8 & 5        &    3.0 & --46$\pm1$      \\
       &      &          &           &               &         &               
&      &      &          &           &     \\
\hline
\hline
\end{tabular}
\normalfont
\smallskip\noindent
\flushleft{\normalsize {
The values for each components are organised as follows: 
column 1, IF number; column 2, observing frequency; 
column 3, total flux density; column 4, peak flux density; 
column 5, 1$\sigma$ rms noise;
column 6, polarised flux density; column 7, 1$\sigma$ rms noise;
column 8, major axis; column 9, minor axis; column 10, major axis PA;
column 11, percentage polarisation; column 12, electric vector
position angle.}}
\end{table*}

The total flux density detected in a combined C-band image
is  1274\,mJy, about 89\% of the flux
density detected by MERLIN (L\"udke et al. 1998) and  85\% of the flux
density detected with the Effelsberg 100-m telescope by
\citet{Mantovani09}. The polarisation percentage is estimated to be
0.96\%, close to the value of 0.8\% measured with MERLIN
\citep{Ludke98}.

\begin{figure*}[t]
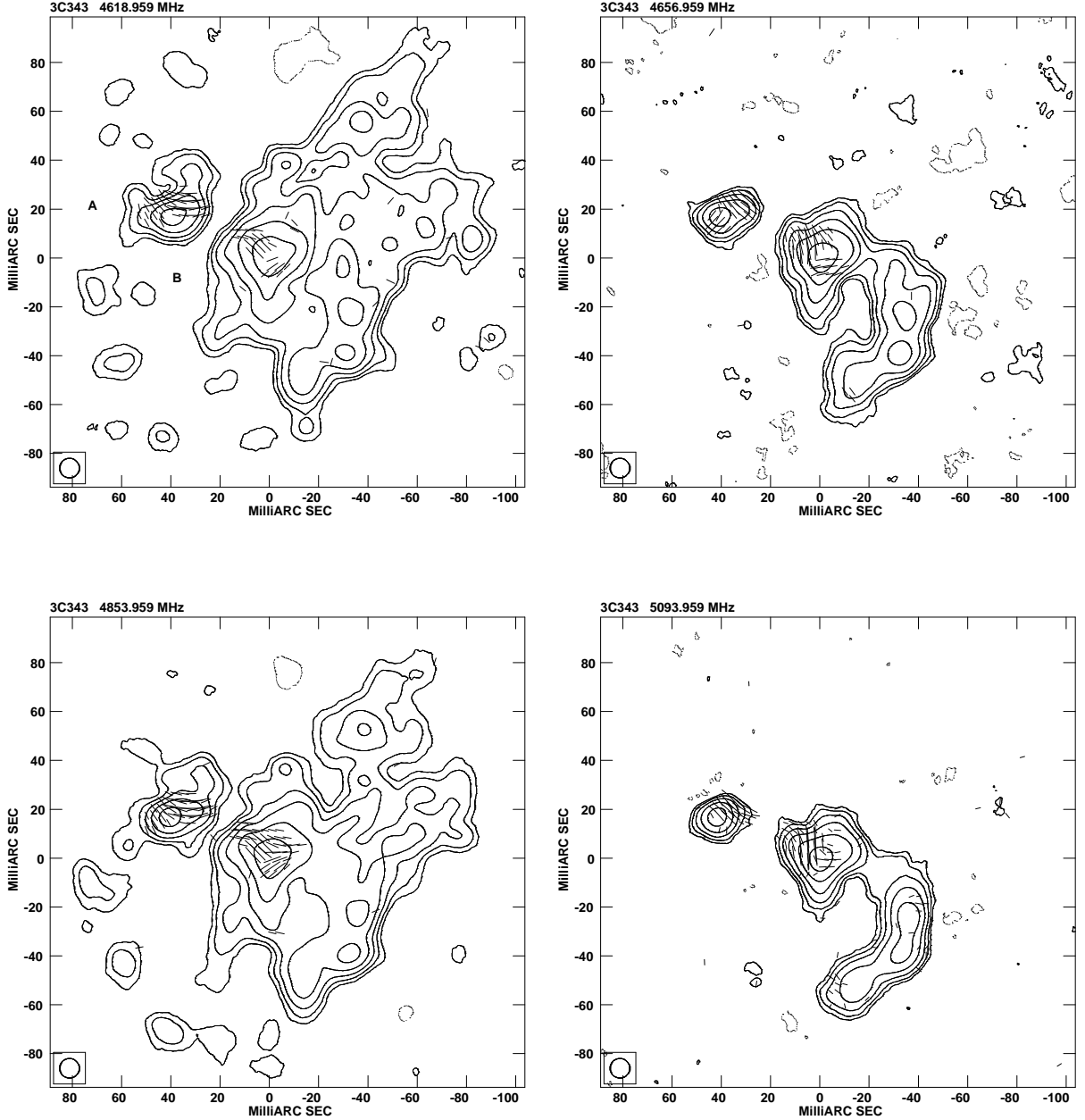

\addtocounter{figure}{+0}
\centering
\includegraphics[width=8cm]{3C343_IF1_F.PS}        
\includegraphics[width=8cm]{3C343_IF2_F.PS}        
\includegraphics[width=8cm]{3C343_IF3_F.PS}        
\includegraphics[width=8cm]{3C343_IF4_F.PS}        
\caption{The total intensity contours for 3C\,343\,A-C 
for the four C-band IFs with the E vectors
superimposed. Contour levels increase by factors of two from 1.5~mJy/beam.
A vector of 1\,mas corresponds to 0.167\,mJy/beam of polarised emission.
The convolution beam is 8$\times$8 mas$^2$.
\label{fig:3c343-Cband}}
\end{figure*}

The total flux density detected in the VLBA image at 8.4\,GHz 
(Fig.~\ref{fig:3c343-Xband}) is 553\,mJy, 69\% of the flux density
detected by \citet{Mantovani09}. The polarised emission detected is of 
7.2\,mJy, corresponding to 1.3\%. The percentage polarisation drops by
a factor of 0.73 between 8.4 and 4.8\,GHz.

%
\begin{figure*}[t]
\addtocounter{figure}{+0}
\centering
\vspace{0.3cm}
\includegraphics[width=8cm,angle=-90]{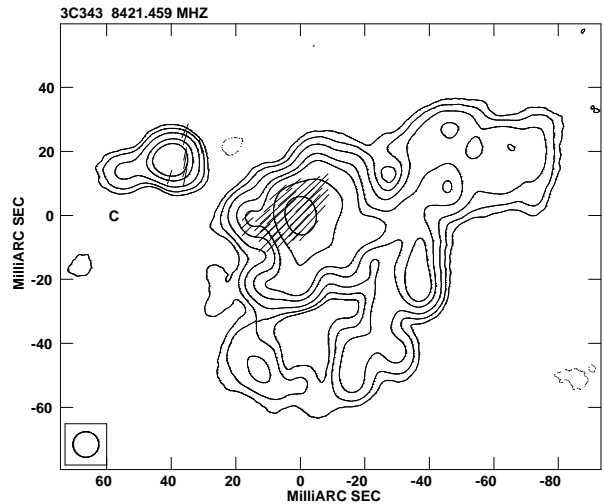}
\vspace{0.3cm}
\caption{The VLBA image of 3C\,343 at 8.4\,GHz using data from the
full bandwidth. Contour levels increase by a factor of two from
0.75 mJy/beam. A vector of 1 mas corresponds to 0.1 mJy/beam of polarised 
emission. The convolution beam is 8$\times$8 mas$^2$.
\label{fig:3c343-Xband}}
\end{figure*}   
\begin{table*}[htbp]
%
\caption{Polarisation parameters of CSSs from mas-resolution observations }
\label{tab:rm-lit}
\begin{tabular}{lccccccll}
\hline
\hline
Name     & Alt. Name  &  z    & Id. &  m                      &  DP                & $RM_{rf}$      &  Notes                           & Ref. \\
         &            &       &     &  (\%)                   &                    & (rad\,m$^{-2}$)&                                  &      \\
\hline
3C\,43     & B0127+233   & 1.459 & Q   & ~0.4$_{\rm{1.6\,GHz}}$  &                    & \,\,\,~14236   &  E, $\sim$1600 pc from core         & 1   \\
3C\,119$^a$& B0429+415   & 1.023 & Q?  & 14.4$_{\rm{8.4\,GHz}}$  & 0.08$_{5.0/8.4}$   &  \,\,\,~10200  &  C, $\sim$325 pc from core          & 2  \\
3C\,138    & B0518+165   & 0.759 & Q   &  ~3.6$_{\rm{5.0\,GHz}}$ &                    &  \,~$-$5287    &  B1, $\sim$40 pc from core          & 3   \\
3C\,147    & B0538+498   & 0.545 & Q   &  ~6.6$_{\rm{8.4\,GHz}}$ &                    & \,~$-$4872     &  BN, $\sim$37 pc from core A$_0$    & 4  \\
4C\,16.14$^b$ & B0548+165& 0.474 & Q   &~6.2$_{\rm{8.4\,GHz}}$   & 0.26$_{4.6/8.4}$   & \,~$-$4275     &  5C, $\sim$540 pc from core         & 5  \\
3C\,216    & B0906+430   & 0.670 & Q   &  30.0$_{\rm{8.4\,GHz}}$ & 0.97$_{4.8/8.4}$   & \,~~~2200      &  Arc, $\sim$1050 pc from core       & 6 \\
3C\,286$^c$& B1328+307   & 0.849 & Q   & 11.0$_{\rm{5.0\,GHz}}$  &                    &                &  Core unclear                       & 7,12   \\
3C\,287$^d$& B1328+254   & 1.055 & Q   & ~4.6$_{\rm{5.0\,GHz}}$  &                    & \,~~~~~$\sim$0 &  Based on only 8.15 and 8.54 GHz    & 8   \\
OQ\,172$^e$& B1442+101   & 3.529 & Q   & ~2.7$_{\rm{8.4\,GHz}}$  &                    &   ~~40000       &  Within $\sim$20 pc from core       & 9,11   \\
3C\,309.1  & B1458+718   & 0.905 & Q   & ~1.5$_{\rm{~15\,GHz}}$  &
0.10$_{8.4/15}$    & \,~$-$1633   & Core {\it RM} between 8.4 and 15 GHz & 10 \\
3C\,318    & B1517+204   & 1.574 & NG  & ~3.5$_{\rm{4.8\,GHz}}$  &                    &  \,~~~3030     &  K2, Core unclear                   & 2   \\
OR$-$140$^f$& B1524$-$136 & 1.687& Q & 30.2$_{\rm{8.4\,GHz}}$  & $\sim$1$_{4.6-5.1/8.4}$   &$-$10000 &  C, $\sim$810 pc from core F        & 5 \\
3C\,343$^g$& B1634+628   & 0.988 & Q   & ~4.8$_{\rm{8.4\,GHz}}$  & 0.91$_{4.6-5.1/8.4}$ & \,\,\,$>$6000     &  core unclear      & 2 \\
3C\,454    & B2249+185   & 1.757 & Q   & 10.9$_{\rm{1.6\,GHz}}$  &                    &  \,~~~5334     & B, $\sim$300 pc from possible core  &   1   \\  
\hline
\hline
\end{tabular}
\smallskip\noindent
\flushleft{\normalsize { Table~\ref{tab:rm-lit} is organised as follows: 
column 1, source name; column 2, alternative name; column 3, redshift; 
column 4, optical identification; 
column 5, polarisation percentage of the emission; 
column 6, highest value of depolarisation;
column 7, highest measured rest frame {\it RM}; 
column 8, notes: letters refer to components designetion as in references
quoted in column 9; column 9, references: 
1) \citet{Cotton03};
2) present work; 3) \citet{Cotton03b}; 4) \citet{Rossetti09};
5) \citet{Mantovani02}; 6) \citet{Venturi99}; 7) \citet{Jiang96};
8) \citet{Dallacasa98}; 9) \citet{Udomprasert97}; 10) \citet{Aaron97};
11) Mantovani, private communication; 12) Cotton et al. (1997) \\
$^a$ m and DP values are for component C, where the peak of polarised 
intensity is surrounded by
a region with {\it RM} in the range of $\sim$9000$-$10200 rad m$^{-2}$;
$^b$ m and DP values are for the entire component 5;
$^c$ fractional polarisations are 8.9 and 11.0\,\% at 5 GHz for 
the western and eastern knots respectively (Cotton et al. 1997);
polarisation is 11.3\,\% for the dominant component with the VLA-A array 
at 4.885 GHz (Perley 1982); 
     {\it RM} expected to be low but not determined with mas resolution;
$^d$ polarisation is 4.6\,\% with the VLA-A array at 4.885 GHz (Perley 1982); 
$^e$ polarisation is 2.0\,\% with the VLA-A array at 4.885 GHz (Perley 1982); 
$^f$ DP estimated using 8.4 GHz and the mean of four measurements between 
4.6 to 5.1 GHz; 
$^g$ DP estimated using 8.4 GHz and the mean of four measurements between 
4.6 to 5.1 GHz for component A;
there appears to be different regions of high {\it RM}. 
}  
}
\end{table*}
We detect polarised emission at all five observing frequencies.
However, due to the low level of polarisation, and to the lower
sensitivity of the observations without the VLA antenna, mapping the
{\it RM} distribution of 3C\,343 is more difficult. It is clear 
from  Fig.~\ref{fig:3c343-Cband} that there could be changes by up 
to $\sim$90$^\circ$ in the orientation of the EVPAs close to the peaks 
of emission in components A and B. Also, significant polarisation is not always 
detected from the same physical region of a component, possibly due
to a combination of both varying sensitivity in the different IFs and
varying {\it RM} across the source. These aspects have made it difficult
to construct a reliable {\it RM} map. However, changes in the EVPAs between
two neighbouring IFs, say IF2 and IF3, by more than $\sim$30$^\circ$ 
in some regions indicate {\it RM} values larger than $\sim$1500 rad m$^{-2}$
in the observer's frame.
While more sensitive observations are required to make a reliable {\it RM}
map, a preliminary one using the present data seems to indicate similarly
high values of {\it RM}. 

\section{Discussion}
\label{sec:discussion}

\subsection{Comments on the three sources}

The three CSSs we present here have mutually very different structures
when imaged with mas resolution. 3C\,119, which is associated with a 
possible quasar, has a complex shape when
observed at low resolution, but has strong core-jet structure on
resolutions of a few mas. We suggest that Component A, which shows a
convex spectrum, flux density variability, and either no, or at best
marginal, polarisation above 
the detection limits, is the core. The jet exhibits ``wiggles''.  From
published images at sub-mas resolution, 3C\,119 appears to maintain
its collimation while presenting an overall  spiral-like shape.
Two bright, polarised blobs are seen along the jet. The brighter of these
has a high rotation measure and strong depolarisation between 8.4 and
5\,GHz, which we interpret as an indication of strong interaction
between the jet and a cloud in the ISM, possibly a dense narrow-line region 
(NLR) cloud.  Although this interaction does not disrupt the jet,
the jet does seem to change direction. The jet could also be bent close
to our line of sight and exhibit apparent superluminal speed. We do not 
detect a counter-jet.
The apparent speed of the two bright blobs (components B and C)
along the approaching jet is likely to be measured by the MOJAVE
monitoring project, and this could possibly put constraints on the orientation 
of the ejection axis.

3C\,318 has an elongated jet, with ``wiggles'' along its southern part.
It is associated with a galaxy. Although the cores in
quasars are often at the end of the jet, this need not be the case for
a galaxy.  From the present observations  we are unable to
identify the exact core. Component K2, suggested by \citet{Spencer91}
 to be the core, has a steep spectrum ($\alpha\approx0.8$) and
shows a percentage polarisation of $\approx$3.4\%  at 4.8\,GHz.
It also has a high rotation measure, often seen in jets in CSS sources.
This component is unlikely to be the radio core.

Among the three sources observed, 3C\,343, which is associated with a
quasar,  presents the most unusual
structure. The two brightest polarised components are embedded in a
diffuse region of weak emission. It is unlikely that either 
is the core of the source. We suggest the component clearly visible 
to the east of component A in the X-band image, instead, as the possible core
candidate. The C-band images without the VLA antenna do not detect the 
extended diffuse emission but show
the two prominent components, A and B, and a ridge of emission which
resembles a strongly curved jet.

The structures of all three CSS sources discussed in this paper 
show large deviations from a collinear structure indicating interactions
of the jets with clouds in the interstellar medium of the host galaxy.  
Although two of the sources, 3C\,119 and 3C\,343, are associated with quasars,
their cores are either weak or undetected, suggesting that projection 
effects due to a small angle of inclination may not be important. However,
jets may be bent towards the line of sight after collisions with clouds.
Systematic monitoring of the knots or peaks of emission may help to clarify
whether this is indeed the case. The detection of high {\it RM}s towards 
the components in the radio jet/structure in both 3C\,119 and 3C\,343, as well
as in 3C\,318, supports the possibility of interaction of the jets with
dense clouds. {\bf For an electron density of $\sim$10$^3$ cm$^{-3}$, which
is a reasonable estimate for the NLR clouds (e.g. Osterbrock 1989; Peterson 1997),
and cloud sizes of $\sim$20 to 100 pc, the highest value of {\it RM} for 
3C\,119 yields magnetic field strengths in the range 
of $\sim$0.1 to 0.6 $\mu$G. The lower value of the cloud size has been taken to 
roughly correspond to the size of the radio components. The corresponding 
values of field strength for 3C\,318 and 3C\,343 are in the range of 
$\sim$0.04$-$0.2 and 0.07$-$0.4 $\mu$G respectively. These values are similar
to those of, for example, Zavala \& Taylor (2002).  For a less dense medium with 
a thermal electron
density of $\sim$1 cm$^{-3}$ and a screen thickness of $\sim$1 kpc, as adopted
by Mantovani et al. (2002), the magnetic field strenghts are in the range of
$\sim$3.7 to 12.6 $\mu$G. These are similar to those estimated by Mantovani 
et al. (2002) for the CSS quasars B0548+165 and B1524$-$136, which also 
have high values of {\it RM} and bends in their radio structures. }   

\subsection{A Discussion of CSS quasars}

To date, few CSSs have been imaged with polarimetric VLBI observations.
Most of those that have are quasars.  At present, 12 out of 24 quasars
in the list of CSSs from the 3C and PW catalogues \citep{Fanti90} have
published polarimetric VLBI data.  Almost all of these show core-jet
structures.  Polarised emission is detected along the jets. Cores are
usually weak and polarisation is not detected, in contrast to flat
spectrum quasars.  CSSs for which  mas-scale values of {\it RM} have
been derived are even rarer.  Table~\ref{tab:rm-lit} summarises
parameters derived from existing observations. Except for the cases of
3C\,287, whose low {\it RM} has been estimated from only two nearby
frequencies, and 3C\,286, which possibly has a low {\it RM} given its
low integrated value but whose {\it RM} has not been determined with
mas resolution, the remaining 12 sources have absolute rest-frame
values of {\it RM} ranging from $\sim$1600 to 4$\times$10$^4$
rad\,m$^{-2}$. In the case of OQ172, which also exhibits the highest
{\it RM}, the region of high {\it RM} is close to the core with the
highest values occurring within $\sim$20 pc of the peak of emission
in the core. This is also the case for the quasar 3C\,309.1, which
has a prominent radio core, with the region of highest {\it RM}
being situated in the vicinity of the radio core.  For the remaining
sources, the component of highest {\it RM} is distinct from the core,
wherever a radio core could be identified. The distance from the core
to the component with the highest {\it RM} varies from $\sim$37 to 1600
pc, with a median separation of $\sim$400 pc. The regions of high {\it
RM} are clearly in the NLR of the host galaxy.

The fractional polarisation usually tends to decrease with decreasing
frequency.  For the sources listed in Table~\ref{tab:rm-lit}, the
components with the highest {\it RM} in 3C\,119 and 3C\,309.1 are
strongly depolarised with DP values of $\sim$0.1 or less (beetween 5.0 and 
8.4\,GHz and 8.4 and 15\,GHz respectively), while those
in 3C\,216 and OR$-$140 (B1524$-$136) exhibit hardly any
depolarisation. While a high {\it RM} without depolarisation could be
due to an external screen, a high {\it RM} along with depolarisation
could be caused by unresolved structures in the screen and/or
thermal plasma mixed with the radio-emitting material. Several examples
of high {\it RM} with little or no significant depolarisation suggests
that the {\it RM} is due to a foreground Faraday screen, with the  NLR
contributing to the observed {\it RM}. In all cases examined, a
$\lambda^2$ law is closely followed over the observed frequency range.
The jets are often distorted and this is interpreted in terms of
jet-cloud interactions, although projection effects can also affect the
observed structure if the jet is bent close to the line of sight. In
many core-jet CSS quasars, high integrated Faraday rotation occurs
where bends in the jet are found, suggesting that jet-cloud
interactions play a significant role in the observed high {\it RM}s of
these components. {\bf For example, amongst the 10 sources in Table 5 
where a core or possible core has been identified, the highest values
of {\it RM} occur at distances from the core ranging from $\sim$20 pc 
for OQ\,172 to $\sim$1600 pc for 3C\,43, excluding 3C\,309.1 where 
the highest value is for the core. The median value of the separation of
the region of highest {\it RM} from the core is $\sim$300 pc. The axes 
of radio emission in these sources bend significantly after the region
of highest {\it RM}. The PA of the latter differs from the axis defined 
by the core and the region of highest {\it RM} by $\sim$20 to 90$^\circ$,
the more extreme cases being 3C\,43, 3C\,119, 4C\,16.14 and 3C\,216. In 
the case of  OQ\,172, the region of highest {\it RM} is within $\sim$20
pc of the nucleus, and the jet shows a large deviation very close to 
the nucleus.}

\subsection{Comparison with pc-scale {\it RM}s in other AGNs}

With the exception of a few radio galaxies having either strong cores
or jets, such as 3C\,111, 3C\,120 and M87, most of the pc-scale
{\it RM} estimates for other AGNs have been made for either
core-dominated quasars or BL Lac objects. Early measurements with
subarcsec resolution using the VLA indicated low core {\it RM}s at long
wavelengths, with the {\it RM}s increasing at shorter wavelengths as
one probes deeper into the radio core (e.g. Saikia et al. 1998, and
references therein). Subsequent mas-resolution observations with the
VLBA have revealed a wealth of information on core {\it RM}s (e.g.
Zavala \& Taylor 2002, 2003, 2004, and references therein; O'Sullivan
\& Gabuzda 2009, and references therein).  A systematic study of the
mas-scale {\it RM} properties of 40 quasars, radio galaxies and BL Lac
objects by Zavala \& Taylor (2003, 2004) has shown that the rest-frame
core {\it RM} for quasars ranges up to $\sim$10$^4$ rad m$^{-2}$ with a
median value of $\sim$1860 rad m$^{-2}$ within $\sim$10 pc of the core.
For BL Lac objects, their core {\it RM} values are usually within
$\sim$1000 rad m$^{-2}$ with a median value of $\sim$440 rad
m$^{-2}$. The {\it RM}s of pc-scale jets decreases rapidly, the
median values of the rest-frame {\it RM} for the jets being $\sim$460
and 260 rad m$^{-2}$ for quasars and BL Lacs respectively (Zavala \&
Taylor 2004).  The few radio galaxies that have been studied exhibit
evidence of moderate to high values of {\it RM}. For example, the core
of 3C\,111 is not significantly polarised, while the jet
exhibits an {\it RM} of $-$750 rad m$^{-2}$~~3\,mas (2.8 pc) east of the
core, decreasing further to $-$200 rad m$^{-2}$~~5\,mas (4.7 pc) east
of the core. For 3C\,120, while Zavala \& Taylor (2002, 2003)
estimate a core {\it RM} of 2080 rad m$^{-2}$, decreasing to
$\sim$100 rad m$^{-2}$ about 1 pc from the core, G\'omez et al. (2008)
find a localised region of high {\it RM} $\approx$3 to 4 mas (2$-$2.6
pc) from the core with a peak {\it RM} of $\sim$6000 rad m$^{-2}$.
The {\it RM} values for M87 could be determined from 18 to 27 mas
($\sim$1.5 to 2 pc) west of the core, and the values varied from
$\sim$$-$5000 to 10$^4$ rad m$^{-2}$ (Zavala \& Taylor 2002, 2003). It
should be noted that M87 is in a cooling core cluster and {\it RM}
values as high as $\sim$8000 rad m$^{-2}$ have been seen towards the
2-kpc radio lobes (Owen, Eilek \& Keel 1990).

In comparison, the median value of {\it RM} for the CSS sources listed
in Table 5 is $\sim$5000 rad m$^{-2}$, and values larger than
$\sim$10000 rad m$^{-2}$ are seen to occur at distances from the core
ranging from $\sim$300 to 1600 pc. This is quite unlike the
regions of {\it RM} discussed earlier, almost all of which are very
close to the radio core. This suggests that in the case of CSS objects,
the jets are often interacting with dense clouds of gas in the
circumnuclear region of the host galaxy.

\subsection{Environmental vs orientation effects}

For quite some time it has been apparent that the degree of core
polarisation correlates with AGN classification. From sub-arcsec scale
measurements with the VLA it was pointed out quite early on that quasar
cores tend to be more polarised than galaxy cores (Saikia, Swarup
\& Kodali 1985; Saikia, Singal \& Cornwell 1987).  Milliarcsec-scale
polarisation measurements also showed a similar trend, with the cores
in BL Lac objects being slightly more polarised than the quasar cores
(Cawthorne et al. 1993). Gabuzda et al. (1992) also presented results
which showed that the cores of BL Lacs are more polarised than
quasars, which was also seen in a larger sample of AGNs by Pollack et
al.  (2003). From arcsec-scale polarisation data, Saikia (1999) showed
that BL Lac objects and core-dominated quasars had higher levels of
core polarisation than lobe-dominated quasars and radio galaxies, and
suggested that this might be due to an orientation effect.  Here, the
low polarisation of the cores of radio galaxies, and perhaps the
lobe-dominated quasars as well, were attributed to depolarisation by
the obscuring torus. However, the observed degree of polarisation could
also reflect the contribution of a small-scale jet, which could be more
strongly polarised and contribute more significantly at smaller
angles to the line of sight.

Taylor (2000) suggested that the {\it RM} values for BL Lac objects
could be smaller than those of quasars, because their jets are believed
to be inclined at smaller angles to the line of sight. This could arise
if the relativistic jets cleared out the magneto-ionic material
responsible for the Faraday rotation.  Similar ideas were suggested by
Saikia et al. (1998) to explain the low {\it RM} values for quasar
cores determined from long-wavelength polarimetric observations with
arcsec resolution. Although individual {\it RM} values in BL Lac
objects are known to exceed thousands of rad m$^{-2}$, the median
value of rest-frame {\it RM} for quasars is larger than that for the BL
Lac objects by a factor of $\sim$4, consistent with the expectation of
Taylor (2000).  It appears that for quasars the long-wavelength
measurements probe the outer regions of the nuclear jets on scales of
tens of pc yielding low {\it RM} values as these are seen through
regions which have been at least partially cleared of the
magneto-ionic medium by the relativistic jets. Shorter wavelength
observations with mas resolution probe deeper into the base of the
jet, and yield high {\it RM}s indicating denser gas and/or higher
magnetic fields than on the larger scales.  It is relevant to note 
that O'Sullivan \& Gabuzda (2009) find the core {\it RM} to
systematically increase with frequency, this being well described by
a power-law, providing information on the power-law fall off of
electron density and/or magnetic field with distance from the nucleus.  
In the scheme where
orientation effects play a role, the cores of radio galaxies and any
emission in their immediate vicinity would be expected to have high
values of {\it RM} due to the effects of the obscuring torus.

{\bf An interesting aspect of our understanding of extragalactic
radio sources is that there are significant differences in the host
galaxy and emission line properties of Fanaroff-Riley class I and II
sources (Fanaroff \& Riley 1974) which may be related to the fuelling
mechanism. While a significant fraction of high-luminosity radio sources
show peculiar optical morphologies and high-excitation emission lines,
reminiscent of gas-rich galaxy mergers, the low-luminosity radio sources
do not share the same optical properties and have weak, low-excitation
emission lines (Heckman et al. 1986; Baum, Heckman \& van Breugel 1992;
Baum, Zirbel \& O'Dea 1995). Hubble Space Telescope observations show
that the low-power radio sources lack evidence of an obscuring torus
and significant emission from a classical accretion disc (Chiaberge, 
Capetti \& Celotti 1999), and may be fuelled by quasi-spherical 
Bondi accretion of circum-galactic gas rather than gas-rich galaxy
mergers (Hardcastle, Evans \& Croston 2007). The finding by Baldi \&
Capetti (2008) that high-excitation radio galaxies almost always show
evidence of star formation, unlike their low-luminosity counterparts,
is consistent with this trend. In our case, BL Lac objects are usually
hosted by galaxies with low-excitation emission lines, while quasars
are hosted in galaxies with high-excitation emission lines. This 
difference is also likely to affect the observed {\it RM}s of BL Lac
objects and quasars, with the former expected to have smaller values.}  

For CSS sources, although orientation is also expected to play a
role (e.g. Saikia et al. 1995), the cores are either weak or not clearly 
identified making it impossible to either determine their {\it RM}s or put 
good limits on their degrees of polarisation.  Amongst the sources listed 
in Table 5, the cores are strong enough for their {\it RM}s to be
estimated in 3C\,309.1 and OQ172, the values being $\sim$$-$1600
and 40000 rad m$^{-2}$ respectively. While 3C\,309.1 exhibits a large
misalignment between the mas-scale and arcsec-scale structures, the jet
in OQ\,172 bends sharply west of the nucleus and the {\it RM} of the jet
falls to less than 100 rad m$^{-2}$ only 10 mas ($\sim$74 pc) from the
nucleus (Udomprasert et al. 1997).  Although the high values of {\it
RM} may be partly due to probing close to the nuclear region, the large
bends in these sources suggests that collisions with  dense
clouds of gas are also a significant factor. For the remaining CSS
objects, where the regions of large {\it RM} occur at distances well
separated from the nucleus, these large values are likely to be due
to interactions of the jets with dense clouds of gas, some of which may
also be fuelling the AGN activity.

\section{Summary and conclusions}
\label{sec:conclusions}
We have presented multi-frequency VLBA polarisation observations of
three CSS sources, namely 3C\,119, 3C\,318 and 3C\,343 to estimate
their {\it RM} values. The radio source 3C\,119 is associated with a
possible quasar, and its $RM$ in the source rest frame has been found
to be as high as $\sim$10200 rad m$^{-2}$ in a region which coincides
with a change in direction of the inner jet.  This component is located
at a projected distance of $\sim$325 pc from the core, which is
almost point-like, variable, has a peaked radio spectrum and is at best
marginally polarised. 3C\,318 is associated with a radio galaxy and its
rest frame {\it RM} has been found to reach a maximum of $\sim$3030
rad m$^{-2}$ for the brightest component which contributes almost all
of the polarised emission. These observations are more sensitive than
those of Spencer et al. (1991) and have detected two more extended
components, which trace  ``wiggles'' in the jet towards the
southern side of the source.  Of the three, the CSS source 3C\,343
has perhaps the most complex structure. It contains two peaks of
emission and a curved jet embedded in more diffuse emission. It
exhibits complex field directions near the peaks of emission, which
indicate rest frame $RM$ values in excess of $\approx$4000 rad
m$^{-2}$.  The varying sensitivity for the different frequencies and
the complex field patterns near the peaks of emission make it
difficult to construct a reliable {\it RM} image for this source.

We have compiled the available data on mas-scale $RM$ estimates for
CSS sources. These show a wide range of values with indications of a
low {\it RM} for 3C\,287 which needs to be confirmed from observations
with a larger number of frequencies, to  values as high as
$\approx$40000 rad m$^{-2}$ in the central region of OQ172
(Udomprasert et al. 1997).  The components with high {\it RM} can also
occur at considerable distances from the core, e.g. in 3C\,43 where
the component with an {\it RM} of $\sim$14000 rad m$^{-2}$ is located
at a projected distance of $\sim$1600 pc (Cotton et al. 2003a).  {\it
RM} estimates for flat-spectrum cores in largely core-dominated radio
sources appear to increase with frequency (see O'Sullivan \& Gabuzda
2009), suggesting that as one probes deeper into the core or unresolved
base of the jet, one samples regions of higher density and/or magnetic
field in the magneto-ionic medium. On larger scales, the jet {\it RM}s
tend to be low as these objects are observed along a line of sight
where the magneto-ionic medium may have been swept out by the
relativistic jets (e.g. Saikia et al. 1998; Taylor 2000; Zavala \&
Taylor 2004).  The CSS objects for which {\it RM} values have been
estimated are almost entirely quasars.  While the effects of an overall
density gradient in the magneto-ionic medium, along with effects of
geometry, orientation {\bf and modes of fuelling of the AGN} , are likely 
to play a significant role, the
high $RM$ values in many of these CSS sources appear to be due to dense
clouds of gas interacting with the radio jets.  Usually, they also
exhibit large structural bends and distortions, consistent with the
possibility of jet-cloud interactions in the interstellar medium of the
host galaxy.  Some of these gas clouds may also be responsible for
fuelling the AGN activity.

\begin{acknowledgements}
{\bf We thank an anonymous referee for his/her very helpful comments
and suggestions, and for a careful reading of the manuscript of this paper.}
The VLBA is operated by the U.S. National Radio Astronomy Observatory
which is a facility of the National Science Foundation operated under
a cooperative agreement by Associated Universities, Inc.
This research has made use of data from the MOJAVE database that is 
maintained by the MOJAVE team (Lister et al., 2009, AJ, 137, 3718).
This research has made use of the NASA/IPAC Extragalactic Database (NED) 
which is operated by the Jet Propulsion Laboratory, California Institute 
of Technology, under contract with the National Aeronautics and Space 
Administration. FM likes to thank Prof. Anton Zensus, Director, for the
kind hospitality at the Max-Planck-Institut f\"{u}r Radioastronomie, Bonn,
for a period during which part of this work has been done.

\end{acknowledgements}

\end{document}